\documentclass{aa}
\usepackage[varg]{txfonts}
\usepackage{subfigure}
\usepackage{rotating}
\usepackage[colorlinks=true, linkcolor=blue, citecolor=blue]{hyperref}
\usepackage{multirow}
\usepackage{float}
\usepackage{ulem}

\begin{document}

\title{The radio spectral  turnover of radio-loud quasars at $z>5$}

\author{Yali Shao\inst{1, 2}\thanks{yshao@mpifr-bonn.mpg.de}
\and Jeff Wagg\inst{3}
\and Ran Wang\inst{2}
\and Emmanuel Momjian\inst{4}
\and Chris L. Carilli\inst{4}
\and Fabian Walter\inst{5, 4}
\and Dominik A. Riechers\inst{6, 5}
\and Huib T. Intema\inst{7, 8}
\and Axel Weiss\inst{1}
\and Andreas Brunthaler\inst{1}
\and Karl M. Menten\inst{1}}

\institute{Max-Planck-Institut f\"ur Radioastronomie, Auf dem H\"{u}gel 69, 53121 Bonn, Germany%1
\and Kavli Institute for Astronomy and Astrophysics, Peking University, Beijing 100871, China%2
\and SKA Observatory, Lower Withington Macclesfield, Cheshire SK11 9FT, UK%3
\and National Radio Astronomy Observatory, Socorro, NM 87801-0387, USA%4
\and Max-Planck-Institut f\"ur Astronomie, K\"onigstuhl 17, D-69117 Heidelberg, Germany%5
\and Department of Astronomy, Cornell University, Space Sciences Building, Ithaca, NY 14853, USA%6
\and Leiden Observatory, Leiden University, Niels Bohrweg 2, 2333 CA Leiden, The Netherlands%7
\and International Centre for Radio Astronomy Research, Curtin University, GPO Box U1987, Perth,
WA 6845, Australia}%8

\abstract{We present Karl G. Jansky Very Large Array (VLA) S- (2--4 GHz), C- (4--8 GHz), and X-band (8--12 GHz) continuum observations toward seven radio-loud quasars at $z>5$. This sample has previously been found to exhibit spectral peaks at observed-frame frequencies above $\sim$1 GHz. We also present  upgraded Giant Metrewave Radio Telescope (uGMRT) band-2 (200 MHz), band-3 (400 MHz), and band-4 (650 MHz) radio continuum observations toward eight radio-loud quasars at $z>5$, selected  from our previous GMRT survey, in order to sample their low-frequency synchrotron emission. Combined with archival radio continuum observations, all ten targets show evidence for spectral turnover. The turnover frequencies are $\sim$1--50 GHz in the rest frame, making these targets gigahertz-peaked-spectrum (GPS) or high-frequency-peaker (HFP) candidates. For the nine well-constrained targets with observations on both sides of the spectral turnover, we fit the entire radio spectrum with absorption models associated with synchrotron self-absorption and free-free absorption (FFA). Our results show that FFA in an external inhomogeneous medium can accurately describe the observed spectra for all nine targets, which may indicate an FFA origin for the radio spectral turnover in our sample. As for the complex spectrum of J114657.79+403708.6 at $z=5.00$ with two spectral peaks, it may be caused by multiple components (i.e., core-jet) and  FFA   by the high-density medium in the nuclear region. However, we cannot rule out the spectral turnover origin of variability. Based on our radio spectral modeling, we calculate the radio loudness $R_{2500\rm\, \AA}$ for our sample, which ranges from 12$^{+1}_{-1}$ to 674$^{+61}_{-51}$.
}


\maketitle

\section{Introduction}
\label{intro} 

Among  the known population of  active galactic nuclei (AGN), about $10\%$ are classified as radio-loud sources (e.g.,  \citealt{Flesch2021}). The origin and evolution of the radio emission remain areas of intense focus in the studies of radio-loud AGN. The intrinsically small radio-loud AGN with radio emission not  generally dominated by their cores have steep radio spectra and are thought to represent the early evolution stage of radio AGN; however, some may be confined to small regions by the ambient interstellar medium (ISM; e.g., \citealt{ODea2021}).

The so-called compact steep spectrum (CSS) sources tend to have projected linear sizes of between 500 pc and 20 kpc and are typically characterized by radio spectral peaks at frequencies below 400 MHz (e.g., \citealt{Fanti1990}). 
The characterized convex synchrotron radio spectra toward gigahertz-peaked-spectrum (GPS) and high-frequency-peaker (HFP) sources peak at about $\sim$1 GHz and $>$5 GHz, respectively (e.g., \citealt{Gopal-Krishna1983}; \citealt{Stanghellini1998}; \citealt{Dallacasa2000}).
The GPS and HFP sources tend to have smaller projected linear sizes of $<$500 pc.
\citet{ODea2021} refer to GPS and HFP sources together as peaked-spectrum (PS) sources. The samples of PS sources may be contaminated by blazars, whose radio spectra can peak at high frequencies due to the behavior of flaring (e.g., \citealt{ODea1983,ODea1986}; \citealt{Kovalev2002, Kovalev2005}). 
\citet{Torniainen2005},  by investigating a sample of mostly quasar-type GPS source candidates with about two decades of data, found that most PS quasars are flat-spectrum sources but have peaked spectra during a flare, and that only a few have a stable convex spectra with low variability.
\citet{Tinti2005} also proposed that quasar-associated PS sources are largely flaring blazars.  Long-term multi-epoch and multifrequency monitoring is necessary to determine whether they are genuine PS sources.

As for CSS and genuine PS sources, the mechanism responsible for the radio spectral turnover will have significant implications on either the internal properties (e.g., magnetic field) or the external environment (e.g., thermal electron density) of the radio source (e.g., \citealt{ODea1998}; \citealt{Tingay2003}; \citealt{ODea2021}). 
The two leading mechanisms for the radio spectral turnover -- synchrotron self-absorption (SSA)
and free-free absorption (FFA) -- may be at work in individual sources.

In the ``young scenario,'' the HFPs are very young radio sources and will develop into extended radio galaxies and quasars (e.g., FR I or FR II) after evolving through the GPS and CSS stages. 
Analytical models with the radio emission from the compact radio sources propagating through their host galaxies provide a reasonable path for compact radio sources evolving into CSS and large-scale radio sources (e.g., \citealt{Begelman1999}; \citealt{Snellen2000}; \citealt{Maciel2014}),  which is in agreement with current observations (e.g., \citealt{An2012}).
The global relationship between radio spectral turnover frequency and the source linear size could be reproduced under  the SSA assumption (e.g., \citealt{Snellen2000}; \citealt{deVries2009}; \citealt{Jeyakumar2016}).
In addition, the magnetic field strength calculated from the turnover, assuming it is produced by SSA for young radio sources, was found to be consistent  with the magnetic field calculated under the assumption that the radio emission is in a near equipartition of energy between the radiating particles and the magnetic field (e.g., \citealt{ODea1998}; \citealt{Orienti2008}).

Under the ``frustration'' hypothesis, the compactness of CSS and PS sources suggests that their radio jets are confined to within small spatial scales by a dense medium in their environments, and, as a consequence, their emission is suppressed (``frustrated'') through the interaction with dense clouds in the host galaxy ISM (\citealt{vanBreugel1984}; \citealt{Peck1999}; \citealt{Tingay2003}; \citealt{Callingham2015}; \citealt{Tingay2015}). This picture is motivated by the fact that CSS and PS sources are much more asymmetric than the large-scale radio sources, which may be caused by interaction with a dense medium (e.g., \citealt{Junor1999};  \citealt{Saikia2001}; \citealt{Thomasson2003}; \citealt{Orienti2007}). Simulation work indicates that jet interaction with large amounts of  dense clouds can disrupt the jets and/or block their propagation (e.g., \citealt{DeYoung1991}; \citealt{Wiita2004}; \citealt{Bicknell2018}). The required gas masses are estimated to be at  least $10^{9}$ to $10^{10}$ $M_{\odot}$ (e.g., \citealt{DeYoung1993}; \citealt{Carvalho1998}; \citealt{ODea1998}). In terms of the molecular gas toward the compact radio sources, there is matching observational evidence for dense clouds in some of their environments (e.g., \citealt{Evans1999}; \citealt{Dasyra2012}; \citealt{Ostorero2017}; \citealt{Morganti2021}). Models with relativistic jet feedback in evolving galaxies can also reproduce the inverse scaling relationship between the peak frequency and the source linear size (e.g., \citealt{Bicknell1997, Bicknell2018}).

High-redshift radio-loud quasars at $z>5$ are critical for probing the physical conditions at the end of the reionization epoch and for studying the early evolutionary stage of  the radio-loud AGN and the coevolution with their host galaxies (e.g., \citealt{Banados2021}; \citealt{Khorunzhev2021};\citealt{Ighina2021}).
Of these objects, four are identified as blazars, which mostly have a core-jet morphology (e.g., \citealt{Romani2004}; \citealt{Frey2010,Frey2015}; \citealt{Spingola2020}).
About half of the objects have been observed at milliarcsecond (mas) resolution by the European Very Long Baseline Interferometry (VLBI) Network (EVN) or the Very Long Baseline Array (VLBA; e.g.,  \citealt{Frey2003,Frey2005,Frey2008,Frey2011}; \citealt{Momjian2003, Momjian2008, Momjian2018, Momjian2021}; \citealt{Romani2004}; \citealt{Cao2014}; \citealt{Gabanyi2015}). 
Various morphologies are observed -- single-compact, double-structure, core-jet,  and dominant core emission with weak unresolved radio extension -- toward these targets. Together with the steep radio spectrum, some are proposed to be compact symmetric objects or  medium symmetric objects. Thus, these high-redshift radio-loud quasars are a unique sample that can be used to shed light on the origin and evolution of radio-loud AGN at the earliest epoch.

\citet{Shao2020} presented Giant Metrewave Radio Telescope (GMRT) 323 MHz radio continuum observations toward 13 radio-loud quasars at $z>5$, detecting their low-frequency synchrotron emission.  Combined with archival radio continuum detections,  nine quasars have power-law spectral energy distributions throughout the radio range. For some, the flux density drops with increasing frequency, with power-law indices from $-0.90$ to $-0.27$ in the frequency range spanning the observed frame  frequency of  a few hundred megahertz to a few gigahertz, while it increases for others, with power-law indices of 0.18 to 0.67 below the observed frame frequency of $\sim$2 GHz.

Multiwavelength observations near the radio spectral peak will allow the determination of the nature of the spectral turnover (FFA vs. SSA) and the relationship between the radio sources and their environments. 
In this paper, to further characterize the entire radio spectral energy distribution of these radio-loud quasars at $z>5$, we report on Karl G. Jansky Very Large Array (VLA) S-, C- and X-band continuum observations toward seven GMRT targets \citep{Shao2020} that have shown or may exhibit spectral peaks at high frequency (e.g., $>1$ GHz) in the observed frame. Taken together, the VLA S, C, and X bands have a continuous frequency coverage from 2 to 12 GHz, which corresponds to 12--72 GHz for an object at $z=5$. The VLA observations can determine if a radio spectral turnover exists at a higher frequency.
We also report on the upgraded GMRT (uGMRT) band-2, band-3, and band-4 radio continuum observations toward eight GMRT quasars that were made to measure their low-frequency synchrotron emission, and we provide a better characterization of their  radio spectral turnover properties (i.e., turnover frequency and the corresponding source strength)  if such turnover exists. 

The outline of this paper is as follows.
In Sect. \ref{obs} we describe our sample, the VLA and uGMRT observations, and the data reduction. 
In Sect. \ref{res} we present the new VLA and uGMRT measurements, apply different spectral models to the observed radio spectra, and measure precise radio loudness based on our radio spectral modeling.
In Sect. \ref{dis}  we discuss the origin of the radio spectral turnover -- variability, FFA, or SSA.
Finally, in Sect. \ref{sum} we present a  summary. 
Throughout the paper we adopt a $\Lambda$ cold dark matter cosmology, with $H_{0}$ = 67.8 km s$^{-1}$ Mpc$^{-1}$, $\Omega_{\rm M}$ = 0.3089, and $\Omega_{\Lambda}$ = 0.6911 \citep{PlanckCollaboration2016},  and a definition of $S_{\nu}\propto\nu^{\alpha}$, where $S_{\nu}$ is the flux density, $\nu$ is the frequency, and $\alpha$ is the spectral index.

\section{Observations and data reduction} 
\label{obs}

We select seven radio-loud quasars from our GMRT project, which observed the 323 MHz radio continuum of 13 radio-loud quasars at $z>5$ \citep{Shao2020}. These objects have shown or may exhibit spectral peaks at frequencies above 1 GHz  in the observed frame (see Figs. 3 and 4 in \citealt{Shao2020}).  The parent radio-loud quasar sample in our GMRT project contains all the radio-loud quasars at $z>5$ identified before 2015, and we also  identified another three by cross-matching optical wavelength quasar catalogs with the VLA Faint Images of the Radio Sky at Twenty-Centimeters (FIRST; \citealt{Helfand2015}) catalog and the VLA high-resolution radio survey of  Sloan Digital Sky Survey (SDSS) Strip 82 \citep{Hodge2011}.
With our VLA S- (2--4 GHz), C- (4--8 GHz), and X-band (8--12 GHz) observations, we investigate whether a spectral turnover exists and locate the turnover position (i.e., frequency and strength).
We also select eight radio-loud quasars from our GMRT project to conduct uGMRT  band-2 (200 MHz), band-3 (400 MHz), and band-4 (650 MHz) observations to well sample the low-frequency radio spectra and to better constrain the turnover property.  In summary, we report new results on ten radio-loud quasars at $z>5$, which are listed in Table \ref{target} of this work, and among them five have both new VLA and uGMRT observations.  
Our sample in this work can represent more than 30$\%$ of the entire radio-loud quasars at $z > 5$ identified by now (e.g., \citealt{Anderson2001}; \citealt{Fan2001}; \citealt{Sharp2001}; \citealt{Romani2004}; \citealt{McGreer2006, McGreer2009}; \citealt{Willott2010}; \citealt{Zeimann2011}; \citealt{Yi2014}; \citealt{Banados2015, Banados2018, Banados2021}; \citealt{Belladitta2020}; \citealt{Gupta2021}; \citealt{Ighina2021}; \citealt{Khorunzhev2021}; \citealt{Liu2021}). 
\begin{table*}
\renewcommand\arraystretch{1.5}
\caption{Targets in this work.}
\label{target}
\centering
\begin{tabular}{lcccccccccc}
\hline\hline
Source&Short name&RA&Dec&$z$&Observatory\\
(1)&(2)&(3)&(4)&(5)&(6)\\
\hline
SDSS J013127.34$-$032100.1&J0131$-$0321&$^{i}$01:31:27.3473&$-$03:21:00.0791&$^{g}$$5.18\pm0.01$&VLA $\&$ uGMRT\\
SDSS J074154.72+252029.6&J0741+2520&$^{d}$07:41:54.72&+25:20:29.6&$^{d}$5.194&VLA $\&$ uGMRT\\
SDSS J083643.85+005453.3&J0836+0054&$^{j}$08:36:43.8606&+00:54:53.232&$^{b}$$5.774\pm0.003$&uGMRT\\
SDSS J091316.56+591921.5&J0913+5919&$^{l}$09:13:16.5472&+59:19:21.6656&$^{e}$$5.1224\pm0.0001$&uGMRT\\
SDSS J103418.65+203300.2&J1034+2033&$^{h}$10:34:18.65&+20:33:00.2&$^{h}$$5.0150\pm0.0005$&VLA $\&$ uGMRT\\
SDSS J114657.79+403708.6&J1146+4037&$^{k}$11:46:57.79043&+40:37:08.6256&$^{e}$$5.0059\pm0.0007$&VLA $\&$ uGMRT\\
FIRST J1427385+331241&J1427+3312&$^{m}$14:27:38.58563&+33:12:41.9252&$^{c}$6.12&uGMRT\\
SDSS J161425.13+464028.9&J1614+4640&$^{h}$16:14:25.13&+46:40:28.9&$^{h}$$5.3131\pm0.0013$&VLA $\&$ uGMRT\\
SDSS J222843.54+011032.2&J2228+0110&$^{n}$22:28:43.52679&+01:10:31.9109&$^{f}$5.95&VLA\\
WFS J224524.2+002414 &J2245+0024&$^{a}$22:45:24.28&+00:24:14.6&$^{a}$5.17&VLA\\
\hline
\end{tabular}
\tablebib{$^{a}$\citet{Sharp2001}; $^{b}$\citet{Stern2003}; $^{c}$\citet{McGreer2006}; $^{d}$\citet{McGreer2009}; $^{e}$\citet{Hewett2010}; $^{f}$\citet{Zeimann2011}; $^{g}$\citet{Yi2014};  $^{h}$SDSS; $^{i}$\citet{Gabanyi2015}; $^{j}$\citet{Frey2003}; $^{k}$\citet{Frey2010}; $^{l}$\citet{Momjian2003}; $^{m}$\citet{Momjian2008}; $^{n}$\citet{Cao2014}}
\tablefoot{Column 1: source name. Column 2: source short name. Columns 3--4: RA and Dec. Column 5: redshift. Column 6: observation facility.}
\end{table*}

\subsection{VLA}

The VLA S-, C-, and X-band radio continuum observations toward these targets were conducted  from 2019 August 07 to September 07 in A-configuration (PI: Yali Shao; Proposal code: 19A-107). The on-source observing times are 116$-$208 s for each target in each band. Flux density scale was established using scans of the standard VLA calibrators: 3C48, 3C138 and 3C147. Details, including information on the  complex gain calibrators, are shown in Table \ref{vlaobs}. 
The total bandwidths are 2 GHz with 16 128 MHz-wide spectral windows at the S band using the eight-bit samplers, and 4 GHz with 32 128 MHz-wide spectral windows at C and X bands using the three-bit samplers.
The data were calibrated using the standard VLA pipeline\footnote{\url{https://science.nrao.edu/facilities/vla/data-processing/pipeline}} in {\ttfamily{CASA}}\footnote{\url{https://casa.nrao.edu/}}. During the data reduction, we also did additional flagging (identification and removal of bad data). 
The final radio continuum images were made using the {\ttfamily{CASA}} task {\ttfamily{TCLEAN}}  with robust weighting factor of 0.5.
For each target, in addition to producing  images of data from all the spectral windows for each observing band (the corresponding results are presented in Table \ref{vlaobs} and Fig. \ref{vlaima}), if the data had high S/N (e.g., > 20) we also imaged each of the  four spectral windows separately in order to characterize the radio spectral slope in each observing band. The measured flux densities are shown as black points in Figs. \ref{nine}--\ref{J1146modelfit}.

\begin{sidewaystable*}
\renewcommand\arraystretch{1.5}
\caption{VLA observations and measurements. }
\label{vlaobs}
%\footnotesize
\centering
\begin{tabular}{lcccccccccc}
\hline\hline
Source &Band&Obs date&$t_{\rm on}$&Complex gain calibrator&Flux density scale calibrator&Beam size&$S_{\rm int}$&$S_{\rm peak}$ \\
&&&(s)&&&(arc sec$^{2}$)&(mJy)&(mJy beam$^{-1}$)\\
(1)&(2)&(3)&(4)&(5)&(6)&(7)&(8)&(9)\\
\hline

J0131$-$0321& S&2019 August 14&116&J0125$-$0005&3C138&$0.74\times0.52$&$60.57\pm0.11$&$59.64\pm0.06$\\
                       &C&2019 August 08&116&J0125$-$0005&3C138&$0.51\times0.29$&$43.34\pm0.11$&$43.15\pm0.06$\\
                       &X&2019 August 07&118&J0125$-$0005&3C138&$0.24\times0.17$&$31.81\pm0.18$&$31.72\pm0.10$\\

J0741+2520 &S&2019 September 07&136&J0741+2706&3C147&$0.58\times0.53$&$3.89\pm0.05$&$3.89\pm0.03$\\
                     &C&2019 August 30&138&J0741+2706&3C147&$0.30\times0.28$&$3.27\pm0.03$&$3.22\pm0.02$\\
                     &X&2019 August 30&208&J0741+2706&3C147&$0.18\times0.17$&$2.46\pm0.02$&$2.46\pm0.01$\\

J1034+2033 &S&2019 September 07&138&J1051+2119&3C147&$0.76\times0.57$&$3.94\pm0.06$&$4.00\pm0.03$\\
                     &C&2019 August 30&136&J1051+2119&3C147&$0.38\times0.29$&$2.92\pm0.04$&$2.85\pm0.02$\\
                     &X&2019 August 30&138&J1051+2119&3C147&$0.26\times0.18$&$1.88\pm0.03$&$1.89\pm0.02$\\

J1146+4037 &S&2019 September 07&128&J1146+3958&3C147&$0.87\times0.55$&$11.05\pm0.06$&$10.76\pm0.03$\\
                     &C&2019 August 30&128&J1146+3958&3C147&$0.43\times0.28$&$9.51\pm0.03$&$9.46\pm0.02$\\
                     &X&2019 August 30&126&J1146+3958&3C147&$0.30\times0.17$&$7.59\pm0.04$&$7.51\pm0.02$\\

J1614+4640 &S&2019 September 02&118&J1613+3412&3C48&$0.90\times0.54$&$4.06\pm0.16$&$3.84\pm0.09$\\
                     &C&2019 August 14&148&J1613+3412&3C48&$0.42\times0.27$&$3.13\pm0.08$&$3.07\pm0.04$\\
                     &X&2019 August 11&148&J1613+3412&3C48&$0.31\times0.19$&$1.96\pm0.10$&$1.80\pm0.05$\\

J2228+0110 & S&2019 September 02&136&J2212+0152&3C48&$0.72\times0.54$&$0.24\pm0.07$&$0.16\pm0.03$\\
                     &C&2019 August 14&196&J2212+0152&3C48&$0.38\times0.27$&$0.22\pm0.03$&$0.21\pm0.02$\\
                     &X&2019 August 11&206&J2212+0152&3C48&$0.22\times0.18$&$0.19\pm0.04$&$0.13\pm0.02$\\

J2245+0024 &S&2019 September 02&138&J2247+0310&3C48&$0.74\times0.54$&$1.65\pm0.06$&$1.61\pm0.03$\\
                     &C&2019 August 14&192&J2247+0310&3C48&$0.38\times0.27$&$1.75\pm0.03$&$1.73\pm0.02$\\
                     &X&2019 August 11&208&J2247+0310&3C48&$0.22\times0.18$&$1.40\pm0.03$&$1.36\pm0.02$\\

\hline
\end{tabular}
\tablefoot{Column 1: source name. Column 2: adopted VLA band. Column 3: observation date. Column 4: on-source integration time. Column 5: complex gain calibrator.  Column 6: flux density scale calibrator. Column 7: the corresponding clean beam size for each observation. Columns 8$-$9: the integrated and peak flux density measured by the {\ttfamily{CASA}} 2D Gaussian tool. Errors shown here are fitting-type errors. During the model fitting in Sect. \ref{rsm}, we also considered additional $5\%$ calibration errors.}
\end{sidewaystable*}

\subsection{uGMRT}

The  uGMRT band 2, 3, and 4 observations were carried out with GMRT Wideband Backend (GWB) from 2020 October 01 to 05 (PI: Yali Shao; Proposal code: 38$\_$069). The on-source observing times are 25 to 71 minutes for each target in each band. We used 3C48 as the flux density scale calibrator for all targets. The details,  including information on  the complex gain calibrators, are shown in Table \ref{ugmrtobs}. Each observing band has one unique spectral window with a bandwidth of 200 MHz.  
We reduced the data using {\ttfamily{CAPTURE}}\footnote{\url{https://github.com/ruta-k/uGMRT-pipeline}} \citep{Kale2020}, which is an automated pipeline to make images from the interferometric data obtained from the uGMRT.  
There was a problem reported concerning an  offset in the calibrated visibilities of baselines between GMRT central square antennas, affecting GWB data taken between 2018 October and 2020 December 03. Baselines between  central square antennas and arm antennas, or between arm antennas, appear to be  affected much less, or not at all. The typical offset  in bands 3 and 4 is $\sim$5--10$\%$ on the central square baselines. So we simply flagged  the 91 baselines between central square antennas (this would affect the root mean square (rms) noise by $\sim$10$\%$). The final radio continuum images shown in Fig. \ref{ugmrtima} were produced by {\ttfamily{CASA}} task {\ttfamily{TCLEAN}}  using a robust weighting factor of 0.5.

\begin{sidewaystable*}
\renewcommand\arraystretch{1.5}
\caption{uGMRT observations and measurements. }
\label{ugmrtobs}
\centering
\begin{tabular}{lcccccccccc}
\hline\hline
Source &Band&Obs date&$t_{\rm on}$&Complex gain calibrator&Flux density scale calibrator&Beam size&$S_{\rm int}$&$S_{\rm peak}$ \\
&&&(min)&&&(arc sec$^{2}$)&(mJy)&(mJy beam$^{-1}$)\\
(1)&(2)&(3)&(4)&(5)&(6)&(7)&(8)&(9)\\
\hline

J0131$-$0321&B2&2020 October 03&36&0116$-$208&3C48&-&-&-\\
                              &B3&2020 October 01&25&0116$-$208&3C48&$14.83\times5.88$&$9.75\pm0.33$&$9.68\pm0.16$\\
                              &B4&2020 October 04&36&0116$-$208&3C48&$6.19\times3.80$&$14.58\pm0.91$&$14.08\pm0.49$\\                              

J0741+2520&B2&2020 October 03&61&0735+331&3C48&-&-&-\\
                           &B4&2020 October 04&36&0735+331&3C48&$4.83\times3.56$&$1.89\pm0.08$&$1.93\pm0.04$\\

J0836+0054&B2&2020 October 03&71&0744$-$064&3C48&-&-&-\\
                           &B4&2020 October 05&36&0744$-$064&3C48&$4.62\times4.03$&$2.06\pm0.28$&$2.67\pm0.19$\\

J0913+5919&B2&2020 October 04&45&0834+555&3C48&-&-&-\\
                           &B4&2020 October 05&36&0834+555&3C48&$5.85\times3.52$&$12.49\pm0.29$&$13.11\pm0.17$\\ 

J1034+2033&B2&2020 October 04&56&3C241&3C48&-&-&-\\
                           &B4&2020 October 05&36&3C241&3C48&$4.51\times3.60$&$4.08\pm0.07$&$4.25\pm0.04$\\

J1146+4037&B2&2020 October 04&40&3C241&3C48&-&-&-\\
                           &B4&2020 October 05&36&3C241&3C48&$5.18\times3.61$&$6.71\pm0.33$&$7.08\pm0.19$\\

J1427+3312&B2&2020 October 04&51&3C286&3C48&-&-&-\\
                           &B4&2020 October 05&36&3C286&3C48&$7.03\times3.64$&$2.29\pm0.25$&$2.65\pm0.15$\\

J1614+4640&B2&2020 October 04&69&3C286&3C48&-&-&-\\
                           &B4&2020 October 05&36&3C286&3C48&$9.38\times3.45$&$1.31\pm0.34$&$1.38\pm0.17$\\

\hline
\end{tabular}
\tablefoot{Column 1: source name. Column 2: adopted uGMRT band. Column 3: observation date. Column 4: on-source integration time. Column 5: complex gain calibrator.  Column 6: flux density scale calibrator. Column 7: clean beam size. Columns 8$-$9: the integrated and peak flux density measured by the {\ttfamily{CASA}} 2D Gaussian tool. Errors shown here are fitting-type errors. During the model fitting in Sect. \ref{rsm}, we also considered additional $10\%$ calibration errors. }
\end{sidewaystable*}

\subsection{Ancillary JCMT archival data}
Target J1146+4037 has two radio spectral peaks, of which the weaker, higher-frequency one may be associated with  thermal dust emission.
In order to constrain the dust emission of J1146+4037, we searched for 850 $\mu$m observations in the   James Clerk Maxwell Telescope (JCMT) archive. We reduced the available data with the {\ttfamily{STARLINK}} SCUBA2 science pipeline\footnote{\url{http://www.starlink.ac.uk/docs/sc21.htx/sc21ch4.html}}. Within the pipeline, we adopted the recipe of {\ttfamily{REDUCE$\_$SCAN$\_$FAINT$\_$POINT$\_$SOURCES}} for detecting extremely low signal-to-noise point sources within blank field images \citep{Chapin2013}. We then used the default flux conversion factor of $537\pm26$ Jy pW$^{-1}$ beam$^{-1}$ \citep{Dempsey2013} and applied the matched filter with a full width at half maximum (FWHM) of 15$\arcsec$.  The reduced 850 $\mu$m map has a noise level of 1.04 mJy beam$^{-1}$. No significant signal was detected toward J1146+4037, and we used a 3$\sigma$ upper limit of 3.12 mJy in the following analysis and discussion.

\section{Results and analysis} 
\label{res}

All  seven VLA targets are detected with  S/N$_{\rm peak}\sim$5$-$1000$\sigma$. The VLA FWHM restoring  beam sizes are $\sim$0$\farcs$6, 0$\farcs$3 and 0$\farcs$2 for the S-, C-, and X-band observations, respectively. The corresponding  rms levels are 0.03--0.09, 0.02--0.06 and 0.01--0.10 mJy beam$^{-1}$ for the VLA S, C, and X band,  respectively.  The details are listed in Table \ref{vlaobs}.

For the eight uGMRT targets, all band-2 observations are contaminated by serious radio frequency interference (RFI). The band-3 and band-4 data are relatively clean, and all sources are detected with S/N$_{\rm peak}\sim$8--106$\sigma$.  The resulted uGMRT synthesized beam sizes are $\sim$6$\arcsec$ and 4$\arcsec$ for bands 3 and 4, respectively. The rms levels are 0.16 and 0.04--0.49 mJy beam$^{-1}$ for bands 3 and 4,  respectively. The details are presented in Table \ref{ugmrtobs}. 

 For both VLA and uGMRT data, we compared the integrated flux density and peak flux density for each target (shown in Tables \ref{vlaobs} and \ref{ugmrtobs}), and they are consistent. In addition, we performed a 2D elliptical Gaussian fit in the image plane using  {\ttfamily{IMFIT}} task in  {\ttfamily{CASA}}, which reports a point source for each target. These indicate that all detected targets are point sources.

We present the distributions of the radio spectra of our sample with new VLA and uGMRT detections (black  points and black open diamonds, respectively) and archival data from the literature (black open squares; \citealt{Frey2003, Frey2005, Frey2008, Frey2010}; \citealt{Momjian2003, Momjian2008}; \citealt{Petric2003};  \citealt{Hodge2011}; \citealt{Cao2014}; \citealt{Coppejans2015, Coppejans2017}; \citealt{Gabanyi2015}; \citealt{Helfand2015}; \citealt{Williams2016};  \citealt{Intema2017}; \citealt{Shao2020}; \citealt{Wolf2021}) in Figs. \ref{nine}--\ref{J1146modelfit}.
The radio spectrum of J2228+0110 shows decreasing flux density with increasing frequency, and at the low-frequency part the upper limit 323 MHz data from \citet{Shao2020} indicates a possible spectral turnover.
The other nine targets show clear evidence for spectral turnover, and among them J1146+4037 shows two spectral peaks.

\subsection{Radio spectral modeling}
\label{rsm}
In order to investigate the turnover origin of the radio spectra of our sample, we fit them with absorption models (see Sect. \ref{amr}). In this work, we assume $5\%$ and $10\%$ calibration errors in addition to the fitting uncertainties presented in Table \ref{vlaobs} and \ref{ugmrtobs} for our new VLA and uGMRT data, respectively, and for the archival data we also added calibration errors. During the model fitting, we used a maximum likelihood method to select the best fitting model, and a MCMC method to better describe the fitting errors using the {\ttfamily{emcee}}\footnote{\url{http://dfm.io/emcee/current/}} package \citep{emcee2013}. The fitting results are presented in Figs. \ref{nine}--\ref{J1146modelfit} and Tables \ref{fitr}--\ref{fitrJ1146}.

\subsubsection{Absorption models}
\label{am}
In this work, we considered two kinds of absorption models to explain the peaked spectra -- SSA and FFA.

{\bf Homogeneous SSA - } The occurrence of the radio spectral turnover in this model is due to the fact that the synchrotron source cannot have a brightness temperature exceeding the plasma temperature of the nonthermal electrons \citep{Kellermann1966}. The inverted radio spectrum cannot have an optically thick slope $>$2.5 under the framework of the typical (i.e., power-law) energy distribution
of the radiating relativistic electrons for  extragalactic radio sources (e.g., \citealt{Slish1963}; \citealt{Pacholczyk1970}). Assuming that the synchrotron source is homogeneous, and a power-law distribution of the electron energy with  power-law index of $\beta$ (e.g., \citealt{Tingay2003}), the spectrum can be described as

\begin{equation}
S_{\nu} = a_{1} \,\bigg(\frac{\nu}{\nu_{\rm p}}\bigg)^{\alpha} \,\bigg(\frac{1-e^{-\tau}}{\tau}\bigg),
\end{equation}where $\alpha$ is the spectral index with $\alpha= -(\beta - 1)/2$, $\tau$ is the optical depth with  $\tau = (\nu/\nu_{\rm p})^{-(\beta+4)/2}$,
$a_{1}$ corresponds to the amplitude of the synchrotron spectrum and $\nu_{\rm p}$ represents the frequency at which the source becomes optically thick.

{\bf FFA - } A distribution of thermal plasma with external or internal ionized-screen relative to the emitting electrons can attenuate/absorb the synchrotron radiation from relativistic electrons.
The morphology of the thermal plasma could be either homogeneous or inhomogeneous in each of the following cases.

Homogeneous FFA case: A standard synchrotron radio source surrounded by  a homogeneous ionized screen will produce the nonthermal power-law spectrum, and the free-free absorbed spectrum can be characterized by 

\begin{equation}
S_{\nu}=a_{2} \,\nu^{\alpha} \,e^{-\tau_{\nu}},
\end{equation}

\noindent where $a_{2}$ and $\alpha$ are the amplitude and the spectral index of the intrinsic synchrotron spectrum, respectively. The optical depth is characterized by $\tau_{\nu} = (\nu/\nu_{\rm p})^{-2.1}$, where $\nu_{\rm p}$ is the frequency at which the optical depth becomes unity (e.g., \citealt{Kellermann1966}; \citealt{Tingay2003}).

Internal FFA case: The absorbing ionized plasma may be mixed in with the relativistic electrons inside the synchrotron source (e.g., \citealt{Kellermann1966}; \citealt{Callingham2015}), as a consequence the peaked spectrum can be parameterized by 

\begin{equation}
S_{\nu}=a_{3} \,\nu^{\alpha} \,\bigg(\frac{1-e^{-\tau_{\nu}}}{\tau_{\nu}}\bigg),
\end{equation}where $a_{3}$ and $\alpha$ are the normalization parameter and the spectral index of the intrinsic synchrotron spectrum, respectively. The optical depth is described by $\tau_{\nu} = (\nu/\nu_{\rm p})^{-2.1}$, where $\nu_{\rm p}$ is at which frequency that the optical depth becomes unity. Below the turnover, the spectral gradient for this model should be $\alpha+2.1$.

Inhomogeneous FFA case: Assuming that the FFA screen is inhomogeneous and external to the synchrotron electrons in the lobes of the source, \citet{Bicknell1997} proposed a bow shock produced by the propagating  AGN jet that will photoionize the ambient ISM. Thus the FFA absorption will occur when the surrounding gas density is significantly high. The model can be reproduced by

\begin{equation}
S_{\nu}=a_{4} \,(p+1) \,\bigg(\frac{\nu}{\nu_{\rm p}}\bigg)^{2.1(p+1)+\alpha} \,\gamma\bigg[p+1, \,\bigg(\frac{\nu}{\nu_{\rm p}}\bigg)^{-2.1}\bigg],
\end{equation}where $a_{4}$, $\alpha$ and $\nu_{\rm p}$ are the normalization parameter, the spectral index and the turnover frequency of the spectrum, respectively,  the term $\gamma\bigg[p+1, \,\bigg(\frac{\nu}{\nu_{\rm p}}\bigg)^{-2.1}\bigg]$ represents the incomplete gamma function of order $p+1$, and $p$ is the distribution of absorbing clouds. The spectral index below the turnover frequency should be $\alpha+2.1(p+1)$.

\subsubsection{Model-fitting results}
\label{amr}
\begin{sidewaysfigure*}
\centering
\includegraphics[scale=0.13]{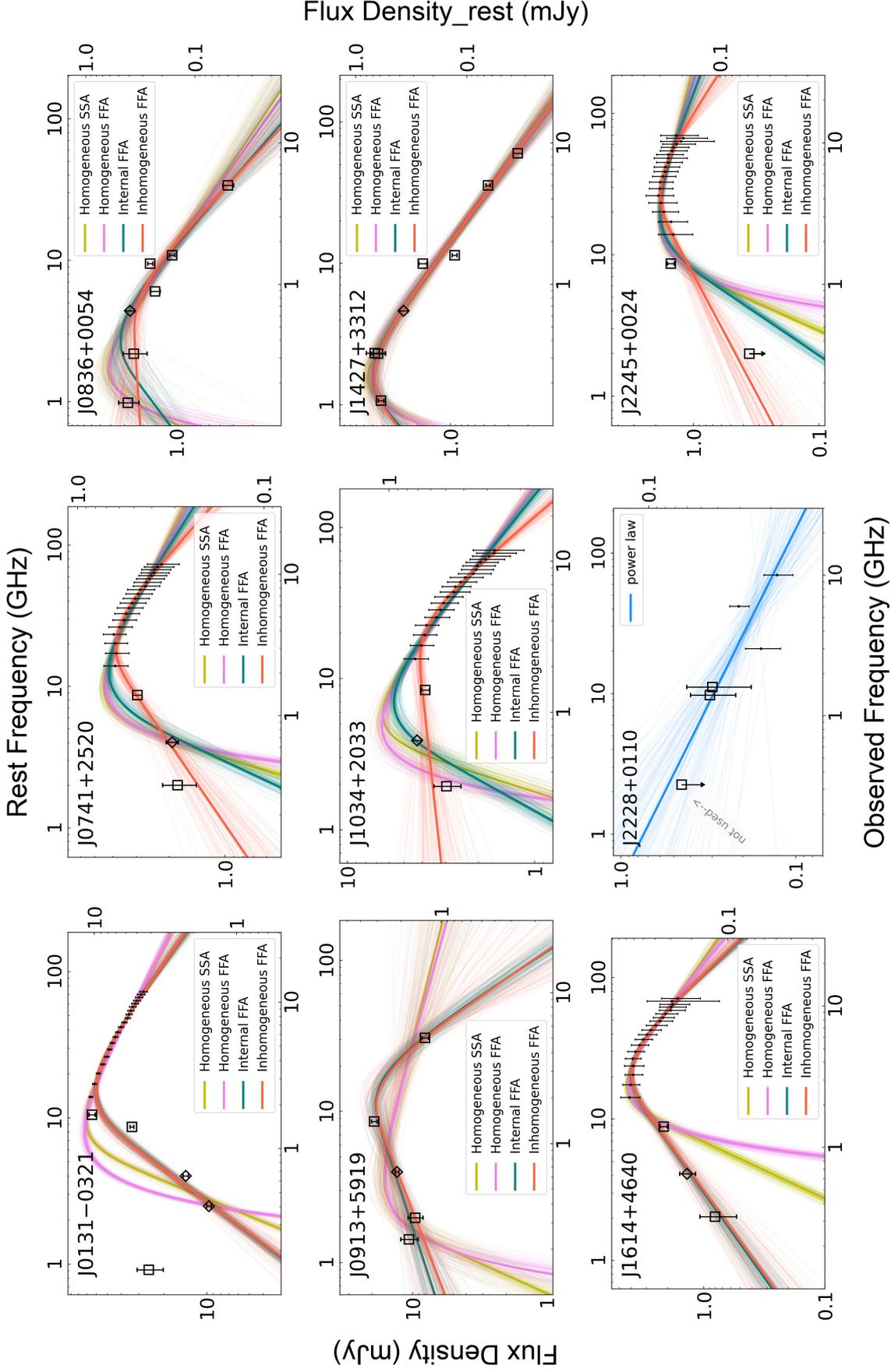}
\caption{Spectral model fit for nine targets in our sample. The black open squares with error bars are archival data taken from the literature. 
The black points with error bars are VLA S-, C-, and X-band measurements. As the errors for the VLA S-, C-, and X-band observations are small, we used five times the uncertainties to make them more visible, except for J2228+0110. We note that we only present the measured flux densities for each observing band toward our targets in Table \ref{vlaobs}. The black open diamonds with error bars are uGMRT measurements, which can be seen in Table \ref{ugmrtobs}. We note that in Tables \ref{vlaobs} and \ref{ugmrtobs} we only list the fitting-type errors for each flux density. In the plot and model fitting we included $5\%$ and $10\%$ calibration errors for our new VLA and uGMRT data, respectively, and for the archival data we also added calibration errors.
For all targets in this figure (except for J2228+0110), we fitted four absorption models -- homogeneous SSA (yellow lines), homogeneous FFA (purple lines), internal FFA (green lines), and inhomogeneous FFA (red lines). The fitted results are presented in Table \ref{fitr}. 
For J2228+0110, the peak signature is not well constrained, and thus we only fit a power-law model (blue lines). And during this model fitting we only used detections, which reveals a flat power law with index of $-0.39^{+0.16}_{-0.17}$.
During the fits, we employed the {\ttfamily{MCMC}} method and visualized the model uncertainties with shaded areas by randomly selecting 100 models from the parameter space. 
 }
\label{nine}
\end{sidewaysfigure*}

\begin{figure*}
\centering
\subfigure{\includegraphics[scale=0.85]{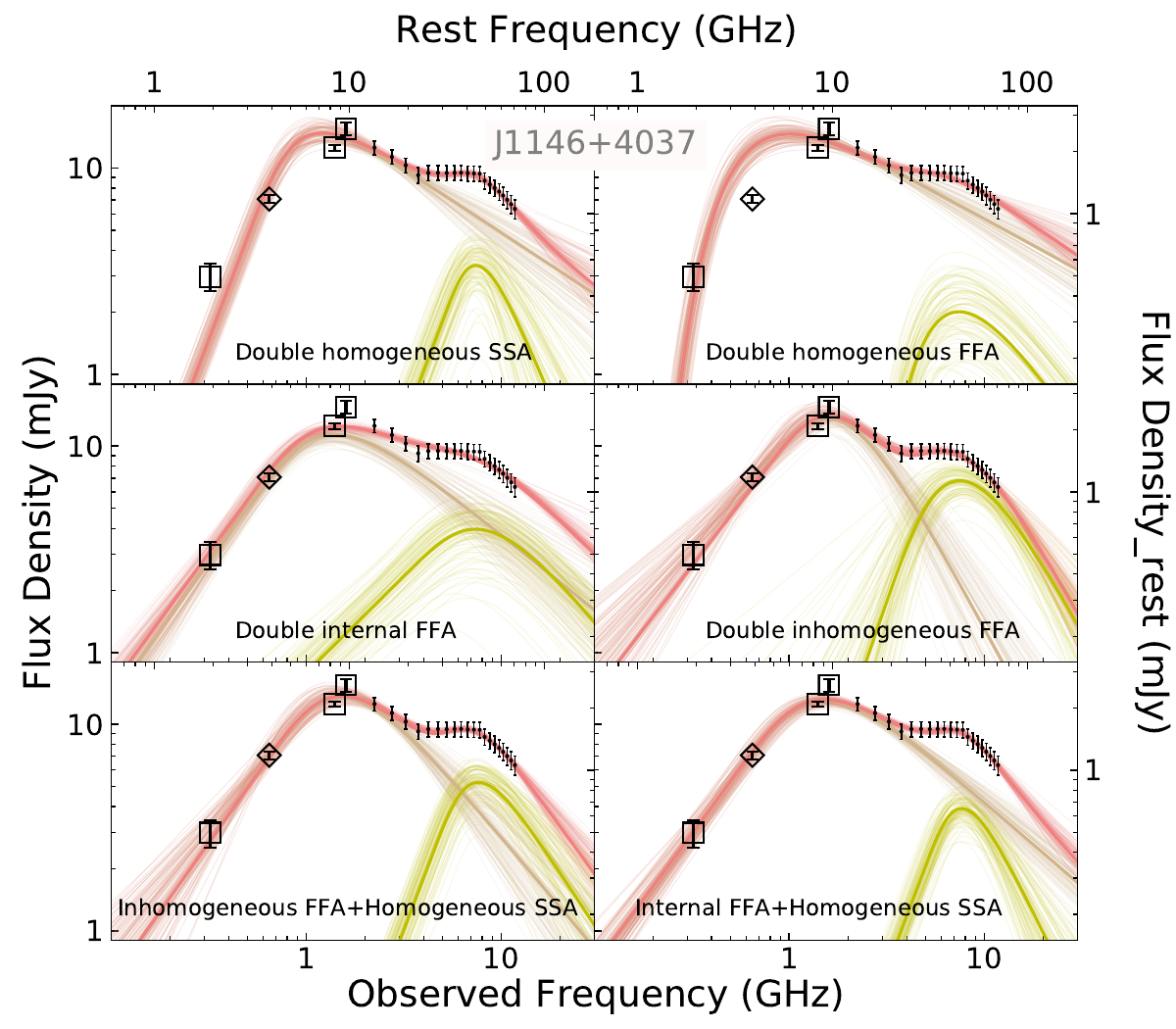}} 
\caption{Spectral model fit with two absorption models for J1146+4037. The symbol description is the same as in Fig. \ref{nine}. The brown and yellow lines represent two absorption models, with names labeled at the bottom of each panel. The red lines are the sum of the two components. The fitted results are presented in Table \ref{fitrJ1146}.  }
\label{J1146modelfit}
\end{figure*}

All ten targets in this paper show evidence for spectral turnover. We fit their entire radio spectra  with all of the above four absorption models and the results are shown in Figs. \ref{nine}--\ref{J1146modelfit} and Tables \ref{fitr}--\ref{fitrJ1146}.  The turnover frequencies are $\sim$1--50 GHz in the rest frame. This makes these targets GPS or HFP candidates.
For J2228+0110, which is less constrained  with a low-frequency upper limit, we only show the fit with a standard nonthermal power-law model: $S_{\nu}\varpropto\nu ^{\alpha}$, where $\alpha$ shows the synchrotron spectral index, and $S_{\nu}$ is the flux density at frequency $\nu$ in GHz, in Fig. \ref{nine}. 
For the well constrained nine targets, the inhomogeneous FFA models can accurately describe their peaked radio spectra. The notes on individual sources are detailed in Appendix \ref{notes}. For J0131--0321, J0913+5919, and J1614+4640, the internal FFA models can also fit their observed spectra well. For J1427+3312, all four absorption models can explain its entire radio spectrum. In particular, for J1146+4037 with two radio spectral peaks, double inhomogeneous FFA models, inhomogeneous FFA+homogeneous SSA models, and internal FFA+homogeneous SSA models can all fit the entire radio spectrum best.

\begin{table*}
\renewcommand\arraystretch{1.8}
\caption{Spectral model fit results for  the one-component case.}
\label{fitr}
\centering
\begin{tabular}{lccccccc}
\hline\hline
 Model&$\alpha$&$\nu_{\rm p}$&$S_{\rm p}$&$\beta$&$p$ &$\alpha_{\rm thick}$&lnlike\\
 &&  (GHz)&(mJy)&& && \\
 (1)&(2)&(3)& (4)&(5)&(6)& (7)&(8)\\
\hline

J0131$-$0321\\
Internal FFA&$-0.74^{+0.04}_{-0.04}$&$13.99^{+0.95}_{-0.92}$&$9.67^{+0.20}_{-0.21}$&-&-        &$1.36^{+0.20}_{-0.21}$&18.17\\
Inhomogeneous FFA&$-0.73^{+0.04}_{-0.04}$&$13.70^{+1.16}_{-1.15}$&$9.66^{+0.23}_{-0.23}$&-&$0.005^{+0.05}_{-0.05}$&$1.38^{+0.12}_{-0.11}$&18.20\\
                       
J0741+2520\\
Inhomogeneous FFA&$-0.68^{+0.07}_{-0.08}$&$20.80^{+3.06}_{-2.66}$&$0.62^{+0.02}_{-0.02}$&-&$-0.43^{+0.05}_{-0.05}$&$0.51^{+0.11}_{-0.09}$&240.59\\                  

J0836+0054\\
Inhomogeneous FFA&$-1.22^{+0.19}_{-0.25}$&$5.90^{+3.39}_{-2.37}$&$0.31^{+0.11}_{-0.08}$&-&$-0.38^{+0.16}_{-0.13}$&$-0.001^{+0.41}_{-0.22}$&18.49\\

J0913+5919\\
Internal FFA&$-1.73^{+0.11}_{-0.10}$&$18.33^{+2.92}_{-2.56}$&$2.48^{+0.31}_{-0.32}$&-&-&$0.37^{+0.11}_{-0.10}$&6.16\\
Inhomogeneous FFA&$-1.67^{+0.57}_{-0.72}$&$17.09^{+5.42}_{-5.36}$&$2.56^{+0.35}_{-0.37}$&-&$0.002^{+0.34}_{-0.26}$&$0.43^{+0.13}_{-0.11}$&5.93\\

J1034+2033\\
Inhomogeneous FFA&$-0.96^{+0.09}_{-0.10}$&$22.89^{+3.69}_{-3.50}$&$0.61^{+0.04}_{-0.04}$&-&$-0.50^{+0.04}_{-0.04}$&$0.08^{+0.08}_{-0.07}$&243.05\\

J1427+3312\\
Homogeneous SSA&$-0.90^{+0.05}_{-0.05}$&$1.32^{+0.10}_{-0.11}$&$0.74^{+0.07}_{-0.06}$&$2.80^{+0.10}_{-0.10}$&-&-&22.52\\
Homogeneous FFA&$-0.93^{+0.05}_{-0.06}$&$1.05^{+0.10}_{-0.11}$&$0.59^{+0.06}_{-0.05}$&-&-&-&22.57\\
Internal FFA&$-0.95^{+0.06}_{-0.06}$&$1.68^{+0.22}_{-0.23}$&$0.69^{+0.05}_{-0.05}$&-&-&$1.15^{+0.06}_{-0.06}$&22.51\\
Inhomogeneous FFA&$-0.94^{+0.06}_{-0.07}$&$1.72^{+0.90}_{-0.37}$&$0.65^{+0.07}_{-0.10}$&-&$-0.09^{+0.66}_{-0.41}$&$0.96^{+1.41}_{-0.88}$&22.51\\

J1614+4640\\
Internal FFA&$-1.27^{+0.07}_{-0.07}$&$26.37^{+1.81}_{-1.75}$&$0.60^{+0.02}_{-0.02}$&-&-&$0.83^{+0.07}_{-0.07}$&231.17\\
Inhomogeneous FFA&$-1.24^{+0.09}_{-0.10}$&$25.76^{+2.28}_{-2.14}$&$0.60^{+0.02}_{-0.02}$&-&$-0.01^{+0.06}_{-0.06}$&$0.83^{+0.14}_{-0.12}$&231.20\\

J2245+0024\\
Inhomogeneous FFA&$-0.75^{+0.14}_{-0.16}$&$29.29^{+5.16}_{-4.65}$&$0.30^{+0.01}_{-0.01}$&-&$-0.36^{+0.07}_{-0.07}$&$0.57^{+0.13}_{-0.10}$&242.02\\

\hline
\end{tabular}
\tablefoot{Column 1:  model case. Column 2: the power-law index of the intrinsic radio source. Column 3: the frequency at which the source becomes optically thick in the rest frame. Column 4: source flux density in the rest frame at the $\nu_{\rm p}$ shown in Column 3. Column 5: the power-law index of the electron energy distribution only for the homogeneous SSA case. Column 6: the distribution of absorbing clouds only for the inhomogeneous FFA case. Column 7: power-law index for the optically thick part (i.e., below the turnover frequency). Column 8: likelihood for the best-fit model. }
\end{table*}

\begin{sidewaystable}
\renewcommand\arraystretch{1.8}
\caption{Spectral model fit results for J1146+4037 with two components.}
\label{fitrJ1146}
\centering
\begin{tabular}{lcccccccccccccc}
\hline\hline
Model&\multicolumn{6}{c}{Low-frequency component}&&\multicolumn{6}{c}{High-frequency component}\\
\cline{2-7}\cline{9-14}
 &$\alpha$&$\nu_{\rm p}$&$S_{\rm p}$&$\beta$&$p$ &$\alpha_{\rm thick}$&&$\alpha$&$\nu_{\rm p}$&$S_{\rm p}$&$\beta$&$p$ &$\alpha_{\rm thick}$&lnlike \\
 & & (GHz)&(mJy)&& && & &(GHz)&(mJy)&& & \\
(1)&(2)&(3)&(4)&(5)&(6)&(7)&&(8)&(9)&(10)&(11)&(12)&(13)&(14) \\
\hline

Double InhFFA&$-1.82^{+0.34}_{-0.42}$&$12.92^{+2.22}_{-2.04}$&$2.04^{+0.22}_{-0.28}$&-&$0.43^{+0.21}_{-0.16}$&$1.15^{+0.27}_{-0.20}$&&$-1.80^{+0.26}_{-0.24}$&$53.00^{+4.53}_{-4.43}$&$1.10^{+0.11}_{-0.13}$&-&$1.17^{+0.61}_{-0.52}$&$2.72^{+1.28}_{-1.00}$&155.76\\

InhFFA+HomSSA&$-1.13^{+0.23}_{-0.24}$&$9.83^{+1.77}_{-1.55}$&$2.24^{+0.14}_{-0.12}$&-&$0.17^{+0.15}_{-0.14}$&$1.30^{+0.28}_{-0.22}$&&$-1.45^{+0.31}_{-0.42}$&$44.09^{+2.73}_{-2.97}$&$0.87^{+0.12}_{-0.16}$&$3.90^{+0.83}_{-0.62}$&-&-&156.38\\

IntFFA+HomSSA&$-0.82^{+0.09}_{-0.09}$&$8.65^{+0.73}_{-0.78}$&$2.17^{+0.08}_{-0.07}$&-&-&$1.28^{+0.08}_{-0.08}$&&$-1.93^{+0.47}_{-0.52}$&$46.52^{+2.54}_{-2.79}$&$1.51^{+0.04}_{-0.04}$&$8.72^{+2.07}_{-1.87}$&-&-&155.80\\

\hline
\end{tabular}
\tablefoot{Column 1: model name. ``InhFFA'' stands for ``inhomogeneous, FFA'' ``IntFFA'' stands for ``internal FFA,'' and ``HomSSA'' stands for ``homogeneous SSA.'' The descriptions for Columns 2--7 and 8--13 are the same as those for Columns 2--7  in Table \ref{fitr}. Column 14: likelihood for the best-fit model.}
\end{sidewaystable}

\subsection{Radio loudness}
By successfully measuring the radio spectral slope around rest frame 5 GHz (see Sect. \ref{rsm}, Figs. \ref{nine}--\ref{J1146modelfit}, and Tables \ref{fitr}--\ref{fitrJ1146}), we are able to precisely calculate the radio loudness defined by \citet{Stocke1992} and \citet{Kellermann1989}:
$R_{2500\rm\, \AA} \ = \ \frac{S_{5\rm\, GHz}}{S_{2500\rm\, \AA}}$
and $R_{4400\rm\, \AA} \ = \ \frac{S_{5\rm\, GHz}}{S_{4400\rm\, \AA}}$, 
where $S_{2500\rm\, \AA}$, $S_{4400\rm\, \AA}$, and $S_{5\rm\, GHz}$ are the rest-frame $2500\rm\, \AA$, $4400\rm \, \AA,$ and 5 GHz flux density, respectively.
We list the rest frame 5 GHz flux densities and the corresponding radio loudness in Table \ref{rl}. The radio loudness $R_{2500\rm\, \AA}$ of our sample, ranges from 12$^{+1}_{-1}$ to 674$^{+61}_{-51}$, which is the first precise radio loudness released for such a large radio-loud quasar sample at $z>5$.
%has SNR $\sim$6--15$\sigma$

\begin{table*}
\renewcommand\arraystretch{1.8}
\caption{Radio loudness.}
\label{rl}
\centering
\begin{tabular}{ccccccc}
\hline\hline
Source&$m_{z}$&$S_{2500\rm\ \AA}$ &$S_{4400\rm\, \AA}$&$S_{5\rm\, GHz}$&$R_{2500\rm\, \AA}$&$R_{4400\rm\, \AA}$  \\
&(mag)&(mJy) &(mJy)&(mJy) &&\\
(1)&(2)&(3) &(4)&(5) &(6)&(7)\\
\hline

J0131$-$0321&$^{c}$18.08&0.046&0.061&$3.83^{+0.28}_{-0.24}$&$83^{+6}_{-5}$&$63^{+5}_{-4}$\\

J0741+2520&$^{c}$18.44&0.033&0.043&$0.38^{+0.03}_{-0.03}$&$12^{+1}_{-1}$&$9^{+1}_{-1}$\\

J0836+0054&$^{c, 1}$18.83&0.022&0.029&$0.33^{+0.03}_{-0.02}$&$15^{+1}_{-1}$&$11^{+1}_{-1}$\\

J0913+5919&$^{c}$20.81&0.0037&0.0049&$2.49^{+0.22}_{-0.19}$&$674^{+61}_{-51}$&$509^{+46}_{-38}$\\

J1034+2033&$^{c}$19.70&0.010&0.014&$0.66^{+0.05}_{-0.05}$&$66^{+5}_{-5}$&$47^{+4}_{-3}$\\

J1146+4037&$^{c}$19.30&0.015&0.020&$1.62^{+0.12}_{-0.10}$&$108^{+8}_{-7}$&$81^{+6}_{-5}$\\

J1427+3312&$^{c, 1}$18.87&0.021&0.027&$0.35^{+0.02}_{-0.02}$&$16^{+1}_{-1}$&$13^{+1}_{-1}$\\

J1614+4640&$^{c}$19.71&0.010&0.013&$0.21^{+0.02}_{-0.02}$&$21^{+2}_{-2}$&$16^{+1}_{-2}$\\

J2228+0110&$^{b, 1}$22.28&0.00089&0.0012&$0.04^{+0.005}_{-0.006}$&$46^{+6}_{-7}$&$34^{+4}_{-5}$\\

J2245+0024&$^{a}$21.72&0.0016&0.0021&$0.15^{+0.03}_{-0.02}$&$94^{+17}_{-11}$&$72^{+13}_{-8}$\\

\hline
\end{tabular}
\tablebib{$^{a}$\citet{Sharp2001}; $^{b}$\citet{Zeimann2011}; $^{c}$SDSS.}
\tablefoot{Column 1: source name. Column 2: $z$-band AB magnitude. Columns 3$-$4: rest-frame 2500 $\rm\AA$ and 4400 $\rm\AA$ flux density, which are calculated assuming  a UV power law $S_{\nu} \propto \nu^{-0.5}$ with $z$-band photometry data. Column 5: rest-frame 5 GHz flux density predicted from our spectral fit. Columns 6$-$7: radio loudness.\\
\tablefoottext{1}{$z$-band magnitude is contaminated by the strong Ly$\alpha$ line emission.}
}
\end{table*}

\section{Discussion} 
\label{dis}

\subsection{Origin of the radio spectral turnover}
Nine of the targets discussed in this work show obvious evidence of spectral turnover, and a possible spectral turnover is implied for another one target considering their radio spectra ranging from 1 to 100 GHz in the rest frame. 
There may be three scenarios to explain the radio spectral turnover toward these nine radio-loud quasars at $z>5$. 
One possibility is that the apparent radio spectral turnover is  due to radio source variability. Other explanations like the model fitting in Sect. \ref{am} are SSA affecting in a small radio-emitting region and  FFA by the dense ambient medium.

\subsubsection{Quasar variability?}
\label{varia}
The entire radio spectra ranging from 1 to 100 GHz in the rest frame presented in Figs. \ref{nine}--\ref{J1146modelfit}, are composed of data taken at different epochs. These observations were taken at times separated by a few hours to a few decades.
Quasars are known to be intrinsically variable, with scales ranging from intraday to much longer time intervals (e.g., \citealt{Jauncey2020}).
The turnover spectra may be an artifact of source variability.  We check source  variability by comparing our VLA S-band observations taken in 2019 with the Very Large Array Sky Survey (VLASS) archival data taken during 2017 and 2021. We find rare variability. However, long-term multifrequency monitoring of the source intensity is still necessary.  
The wide bandwidth (2--4 GHz; corresponding to 12--24 GHz at $z=5.0$) and the multi-spectral-window setup of the VLA S, C, and X bands allow us to characterize the radio spectral change in each observing band. 
For three of them (J0741+2520, J1146+4037, and J2245+0024), we are observing the spectral peaks  inside one VLA band.
However, it is still difficult to determine if we measure the real spectral slope without having long-term multifrequency monitoring.
As variability may cause an apparent spectral slope when using data at different frequencies observed at different times (e.g., \citealt{Tinti2005}).
In our analysis in Sect. \ref{am}, we assume the turnover is real and apply absorption models.

\subsubsection{SSA or FFA?}
\label{ssaffa}
In a young scenario, as the source grows, the inner region (possibly a tiny radio lobe) expands, and as a result, the turnover frequency moves to lower values. In this scenario, the HFPs are newborn radio sources that develop into extended radio sources (e.g., FR I or FR II) after evolving through the GPS and CSS stages. 
Under the SSA assumption, the magnetic field $H$ can be measured directly from the spectral peak parameters - the peak frequency $\nu_{\rm p}$ in units of GHz in the observed frame, and the corresponding flux density $S_{\rm p}$  in units of Jy and source angular sizes $\theta_{\rm maj}$ and $\theta_{\rm min}$ in units of mas at the turnover: 
$H \sim f(\alpha)^{-5} \theta_{\rm maj}^{2} \theta_{\rm min}^{2} \nu _{\rm p}^{5} S_{\rm p}^{-2} (1 + z)^{-1}$. $f(\alpha)$ weakly depends on $\alpha$, and a value of 8 when $\alpha = -0.5$ is usually adopted \citep{Kellermann1981}. The difficulties are determining the turnover frequency and measuring the source size at the turnover frequency. One may test this hypothesis by comparing the magnetic field strength mentioned above with the equipartition magnetic field strength, which can be determined from Eqs. 1--3 in \citet{Orienti2012}.
For example, \citet{Orienti2008} reported agreement in the magnetic field strength measured using the observed turnover information and assuming  equipartition by investigating five HFPs at $0.084<z<1.887$.

Another popular explanation for the turnover and compact natures of CSSs, GPSs, and HFPs is the frustration hypothesis. This theory argues that these sources are confined within small spatial scale and high-density environments, and as a consequence the radio emission is frustrated by the abundant nuclear plasma (\citealt{vanBreugel1984}; \citealt{Peck1999}; \citealt{Tingay2003}; \citealt{Callingham2015}; \citealt{Tingay2015}). High-resolution, ISM observations of the nuclear region of these targets may address this issue. Dense clouds in the environments of some compact radio sources have been detected (e.g., \citealt{Dasyra2012}; \citealt{Ostorero2017}; \citealt{Morganti2021}) with gas masses of $10^{9}$ to $10^{10}$ $M_{\odot}$. Absorption model fits to individual radio sources also favor FFA over SSA (e.g., \citealt{Callingham2015}; \citealt{Mhaskey2019}).
 
Nine sources in our sample have more than one detection on both sides of the spectral turnover. 
In Sect. \ref{am} we implemented a series of models to fit the data for the case of SSA in a homogeneous source, the case of FFA in an internal homogeneous medium, the case of FFA in an external homogeneous medium, and the case of FFA in an external inhomogeneous medium. For J1146+4037, two spectral peaks are detected, which may be caused by multiple radio components. We implemented a collection of any two of the above four cases (repeating one case is also a solution) to fit the observed spectrum toward this target.  The fitting results can be seen in Tables \ref{fitr} and \ref{fitrJ1146}, where we list the likelihood values for the best fitting models. For targets J0741+2520, J0836+0054, J1034+2033, and J2245+0024, we only present the likelihood values of the inhomogeneous FFA models as they are  obviously higher than those of other three models. For targets J0131--0321, J0913+5919, and J1614+4640, the likelihood values of the inhomogeneous FFA models and the internal FFA models are comparable with each other. For target J1427+3312, the likelihood values of all four absorption models are similar, as the optically thick part spectrum is less constrained. For target J1146+4037, the likelihood value of the double inhomogeneous FFA model is better than that of the double homogeneous SSA, double homogeneous FFA, and double internal FFA models.
In summary, the inhomogeneous FFA case can accurately describe all of our targets (double inhomogeneous FFA  for J1146+4037). This means that the radio spectral turnover of GPS and HFP candidates in our work all suggest an FFA origin.

In addition, we can investigate the spectral turnover origin from the aspect of the magnetic field strength as mentioned above. 
Five (J0131$-$0321, J0836+0054, J0913+5919, J1146+4037, and J1427+3312) of nine sources with obvious evidence of spectral turnover have milliarcsecond  observations by the VLBA or the EVN (e.g., \citealt{Momjian2003, Momjian2008}; \citealt{Frey2005, Frey2010}; \citealt{Gabanyi2015}). 
Only two targets (J0131$-$0321 and J1146+4037) have source size measurements with error bars.
With the generic curved model defined by Eq. 1 in \citet{Snellen1998}, which only permits a fit to the spectra on regions at the lower and higher frequencies around the peak but cannot discriminate the underlying absorption mechanism (e.g., SSA or FFA) causing the spectral turnover, we get a turnover frequency of $13.57^{+1.26}_{-1.28}$ GHz in the rest frame and peak flux density of $9.67^{+0.23}_{-0.21}$ mJy in the rest frame for J0131$-$0321.  
And with double generic curved models, we get turnover frequencies of $8.54^{+0.10}_{-0.10}$ and $49.57^{+0.11}_{-0.13}$ GHz in the rest frame, peak flux densities of $2.26^{+0.15}_{-0.13}$ and $0.89^{+0.15}_{-0.16}$ mJy in the rest frame for the two-peaked features in J1146+4037, respectively.

\citet{Gabanyi2015} present a source size of $0.79\pm0.01$ mas  for J0131$-$0321 at an observed frequency of 1.7 GHz obtained with  EVN observations.  The magnetic field strength calculated from the turnover information under the assumption of SSA for J0131$-$0321 is $289^{+135}_{-137}$ mG, which is about four times higher than the equipartition magnetic field strength ($77^{+16}_{-15}$ mG). This may indicate that  the turnover is not caused by  SSA. However, the magnetic field strength calculated from the turnover information under the assumption of SSA only has $2\sigma$ significance. 
\citet{Frey2010}  present a source size of $0.74\pm0.01$ mas  for J1146+4037 at observed frequency of 5 GHz also based on EVN observations.  The magnetic field strength calculated from the turnover information under the assumption of SSA for the $\sim$10 GHz peak is $505^{+75}_{-67}$ mG, which is about one order of magnitude higher than the equipartition magnetic field strength ($56^{+12}_{-11}$ mG).  This may indicate that the turnover is not caused by  SSA. 
These are consistent with our spectral model fitting in Sect. \ref{am} for these two targets.

The magnetic field strength calculated from the turnover information under the assumption of SSA for $\sim$50 GHz peak of J1146+4037 is $\gg$1 G, which is far larger than the equipartition magnetic field strength ($56^{+12}_{-11}$ mG). This may indicate that SSA is not sufficient to explain the $\sim$50 GHz peak. 
Thus, the model combinations of inhomogeneous FFA+homogeneous SSA  and internal FFA+homogeneous SSA presented in Fig. \ref{J1146modelfit} and Sect. \ref{am} may not be applicable. And the double inhomogeneous FFA model shown in Fig. \ref{J1146modelfit} is a much better explanation for the observed double radio peaks.
We note that we use a source size measured at an observed frequency of 5 GHz, but not at the turnover frequency ($\sim$50 GHz in the rest frame corresponding to $8$ GHz in the observed frame at $z=5$). Therefore, this may introduce a very large uncertainty.  For the remaining turnover sources in our sample, we have an ongoing VLBA project to measure the source size at the spectral turnover frequencies, in order to investigate their spectral turnover origin, and also to study the milliarcsecond-scale source structure.

\subsubsection{Physical properties of the absorbing medium}

As presented in Sect. \ref{res}, nine targets in our sample show obvious evidence of spectral turnover. And all of them can be accurately fit by the inhomogeneous FFA model parameterized by \citet{Bicknell1997}. From their advanced simulations of relativistic jets interacting with a warm, inhomogeneous medium, \citet{Bicknell2018} proposed that GPS and CSS radio sources are the signposts of the feedback of relativistic jets on their host galaxies. 
They assume the density $n$ of the ionized medium (ions plus electrons) follows a log normal distribution $\ln n$, which is appropriate for a turbulent medium. 
The FFA of clumpy gas ionized by the radio jets is able to account for the spectral turnover at GHz frequencies as well as the low-frequency power laws observed for GPS and CSS radio sources. The radio spectra of our targets turn over at $\sim$1--50 GHz, with most peaking at >10 GHz in the rest frame. Even through most of our peaked sources are HFPs candidates, we still use the analytical models given by \citet{Bicknell2018} to measure the densities  of the clumpy ISM.
The log normal distribution $\ln n$ has the probability distribution function
\begin{equation}
P(n)=\frac{1}{ns\sqrt{2\pi}}\exp\bigg[-\frac{(\ln n-m)^{2}}{2s^{2}}\bigg], 
\end{equation}
where $m$ is the mean log density and $s$ is the width of the distribution in log density. Then the expected value $E(n)$ of the density (the mean density $\mu$) and the variance $\sigma^{2}$ of the density distribution are $\mu=E(n)=e^{m+\frac{1}{2}s^{2}}$ and $\sigma^{2}=\mu^{2}(e^{s^{2}}-1),$ respectively. When solving $\mu$ and $\sigma^{2}$, following \citet{Zovaro2019} we assume that the only ions contributing to FFA are H$^{+}$, He$^{+}$, and He$^{++}$ and that the ionized medium fractional abundances are $n_{e}/n=0.47175$, $n_{\rm H^{+}}/n=0.41932$, $n_{\rm He^{+}}/n=0.024458,$ and $n_{\rm He^{++}}/n=0.013770,$ respectively, for electron and ions.
Also assuming a typical Baryonic velocity dispersion of 250 km s$^{-1}$ (e.g., \citealt{ForsterSchreiber2009}) and the depth of the absorbing screen to be the same with the source size of 5 pc (available EVN observations toward J0131--0321 and J1146+4037 reveal source sizes around $\sim$0.8 mas corresponding to $\sim$5 pc; \citealt{Frey2010}; \citealt{Gabanyi2015}; we assume each component of J1146+4037 has a size of 2.5 pc) and using Eqs. 10--18 in \citet{Zovaro2019} and Eqs. C1--C2 in \citet{Bicknell2018}, we derive the $\mu$ of the ionized medium to be $3750^{+335}_{-330}$,  $4855^{+815}_{-743}$, and $11075^{+12151}_{-6569}$ cm$^{-3}$ for J0131$-$0321 and the two components of J1146+4037, respectively. The corresponding $\sigma^{2}$ are $5.62^{+1.05}_{-0.95} \times 10^{8}$, $9.43^{+3.43}_{-2.67} \times 10^{8}$, and $5.32^{+16.30}_{-4.51} \times 10^{9}$. The densities for the two components in J1146+4037 are consistent within  their large uncertainties.  Our densities are larger than those (a few tens of cm$^{-3}$) commonly inferred for CSS or GPS sources presented by, for example, \citet{Zovaro2019} and \citet{Mhaskey2019}. We note that the mean density $\mu$ is sensitive to turnover frequency and the depth of the absorbing screen.

Another way to derive the density is making use of emission measure (EM) assuming the electron number density is uniform for a homogeneous absorbing medium. Using the formula provided in \citet{Mezger1967} for a homogeneous \ion{H}{II} region, the optical depth can be expressed as a function of frequency $\nu$, electron temperature $T_{\rm e}$, and EM: 
\begin{equation}
\tau_{\nu} = 0.082\times\bigg[\frac{T_{\rm e}}{{\rm K}}\bigg]^{-1.35}\times\bigg[\frac{\nu}{{\rm GHz}}\bigg]^{-2.1}\times\bigg[\frac{{\rm EM}}{{\rm pc\,cm^{-6}}}\bigg].
\end{equation}
Thus, with the derived turnover frequencies, where $\tau_{\nu}=1$, and assuming $T_{\rm e}$ of $10^{4}$ K, we derive EMs of  $7.47^{+1.39}_{-1.25}\times10^{8}$, $6.27^{+2.27}_{-1.77}\times10^{8}$, and $3.49^{+9.95}_{-2.95}\times10^{9}$ pc cm$^{-6}$ for J0131$-$0321 and the two components of J1146+4037, respectively.
Emission measure is a function of the average electron number density ($n_{\rm e}$) and path length ($L$) through plasma:  EM = $n_{\rm e}^{2}\times L$.
Assuming the path length ($L$) is the same with the source size (we assume each component of J1146+4037 has a size of 2.5 pc), we calculate the electron number density $n_{\rm e}$ of $1.22^{+0.11}_{-0.11}\times10^{4}$, $1.58^{+0.26}_{-0.24}\times10^{4}$, and $3.59^{+3.88}_{-2.12}\times10^{4}$ cm$^{-3}$  for J0131$-$0321 and the two components of J1146+4037, respectively.
The derived electron number densities are consistent with that (a mean $n_{\rm e}$ of $1.6\times10^{4}$ cm$^{-3}$) of a sample of 114 young \ion{H}{II} regions in the Galaxy \citep{Yang2021}. 
The $n_{\rm e}$ can be up to $10^{5}$ cm$^{-3}$ for  hyper-compact \ion{H}{II} regions (e.g., \citealt{MurphyTara2010}).
Assuming an electron fractional abundance that is the same as the above inhomogeneous case of $n_{e}/n = 0.47175$, we get the uniform density of the ionized medium of  $2.59^{+0.23}_{-0.23}\times10^{4}$, $3.36^{+0.56}_{-0.51}\times10^{4}$, and $7.61^{+8.23}_{-4.50}\times10^{4}$ cm$^{-3}$  for J0131$-$0321 and the two components of J1146+4037, respectively.

These two densities are based on different distribution assumptions. We prefer the former calculations  as our fitting results suggest an ``inhomogeneous FFA case,'' which assumes that the FFA screen is inhomogeneous and external to the synchrotron electrons in the lobes of the source. 
Taking J0131$-$0321 as an example, if we compare these two densities,  the uniform $n$ of $2.59^{+0.23}_{-0.23}\times10^{4}$ cm$^{-3}$  is inside the 1$\sigma$ range of the mean density $\mu$ of [0, $2.75\times10^{4}$] cm$^{-3}$, but at the high end.
Other physical parameter such as the ionized pressure ($P_{i}$) can be further measured. 
We calculate the ionized pressure ($P_{i}=n_{\rm e}kT_{\rm e}$, where $k$ is Boltzmann's constant), which are $2.4\times10^{-9}$, $3.2\times10^{-9}$, and $7.2\times10^{-9}$ dyne cm$^{-2}$ for J0131$-$0321 and the two components of J1146+4037, respectively, using the mean density. 
We also calculate the  minimum  pressure $p_{\rm min}$, which is about (1/3)$u_{\rm min}$, where $u_{\rm min}$ is the minimum energy density, and assuming an ellipsoidal geometry and a filling factor of unity (which means that the source is fully and homogeneously filled with the relativistic plasma; e.g., \citealt{Pacholczyk1970}). The minimum  pressure $p_{\rm min}$ are $1.8\times10^{-4}$ and $1.0\times10^{-4}$ dyne cm$^{-2}$ for J0131$-$0321 and J1146+4037, respectively, which are within the typical $p_{\rm min}$ values with a range of $\sim10^{-4}-10^{-6}$ dyne cm$^{-2}$ for a sample of 17 faint HFPs at $0.2<z<2.9$ \citep{Orienti2012}. However, the minimum  pressure $p_{\rm min}$ of our two targets are higher than their ionized pressure ($\sim10^{-9}$ dyne cm$^{-2}$), as well as the minimum  pressure $p_{\rm min}$ ($\sim10^{-9}$, $\sim10^{-11}$--$10^{-10}$ and $\sim10^{-10}$ dyne cm$^{-2}$, respectively) of the hotspots, lobes, and kiloparsec-scale radio jet  in the ultra-luminous radio galaxy Cygnus A \citep{Carilli1996}. We should note that the minimum energy pressure uncertainties are firmly related to the assumptions of the filling factor,  which may differ by several orders of magnitude \citep{Orienti2012}.

\subsection{The spectrum of J1146+4037}
\label{cs}

In addition to the spectral turnover at rest frame $\sim$10 GHz of J1146+4037, we detect an extra weaker peak at $\sim$50 GHz in the rest frame, which spans a frequency range of 20--70 GHz. The significance of the detection of the weaker peak at $\sim$50 GHz is about 10$\sigma$, which is roughly estimated as the S/N of its modeled peak flux density (see Table \ref{fitrJ1146}) within the double inhomogeneous FFA model. 
To exclude the possibility of the rest frame $\sim$50 GHz peak being an artificial product due to calibration, we checked the spectra of both the complex gain calibrator  - J1146+3958 and the flux density scale calibrator - 3C147 used for VLA  observations for J1146+4037. Like for our target J1146+4037, we imaged the data of the calibrators using every four spectral windows so that we have enough data to observe the trend of their spectra. Both calibrators show smooth power-law radio spectra. Thus we can discount the possibility that the $\sim$50 GHz peak is an artifact.

\subsubsection{The $\sim$50 GHz peak - A star formation related component?}
We next investigate the contributions to the $\sim$20--70 GHz peak. There are four possible components: thermal dust emission, which is mainly from dust heated by ultraviolet (UV) radiation from young and massive stars in the host galaxies; thermal free-free radio emission, which is directly related to production rate of ionizing photos in massive (i.e., $\ge$$8\,M_{\odot}$) star forming \ion{H}{II} regions;   nonthermal synchrotron radio emission, which arises from cosmic-ray  electrons and positrons propagating through the magnetized ISM after having been accelerated by supernova remnants; nonthermal synchrotron radio emission, which is associated with central AGN.

As for the thermal dust emission, \citet{Leipski2014} presented the Herschel far-infrared (FIR) wavelength observations toward our target J1146+4037. However, it is undetected. We searched for this target in the JCMT archive, and the data reveal a non-detection with a 3$\sigma$ upper limit of 3.12 mJy at observed frame 850 $\mu$m.  Assuming a dust temperature of 47 K (the average value of the high-redshift quasar sample in \citealt{Beelen2006}), and an emissivity index of 1.6, we predict an upper limit of $5\times10^{12}L_{\odot}$ for the infrared (IR; 8--1000 $\mu$m) luminosity, and the contribution of the thermal dust being $\sim10^{-6}$ of the 50 GHz rest frame emission.

As for the thermal free-free radio emission and nonthermal radio synchrotron emission associated with massive (i.e., $\ge$$8\,M_{\odot}$) star formation, we predict that the combined flux density would contribute $\sim10^{-4}$ of the 50 GHz rest-frame emission, using the predicted IR luminosity upper limit and Eq. 10 in \citet{Murphy2012}. The $\sim10^{-4}$ fraction is also true for the frequency range of 20--70 GHz. The portion between the thermal free-free radio emission and nonthermal synchrotron radio emission  associated with massive star formation at 50 GHz is 3:1. Thus, we can ignore the contribution of these two components. 
In addition, we can also ignore these two emission components in terms of the spectral indices of the $\sim$50 GHz peak. The thermal free-free emission normally exhibits an optically thin flat spectrum with power-law index of $-0.1$ and an optically thick steep spectrum with power-law index of 2, and the nonthermal synchrotron component has a steep spectrum with index of $-$0.8 (e.g., \citealt{Condon1992}). We model the radio spectrum of J1146+4037 with two generic curved model, and get an optically thin power-law index of $-1.83$ and an optically thick power-law index of 1.72 for the $\sim$50 GHz peak. Both spectral indices are too steep for a thermal free-free emission. 
It is unlikely that the formation of the $\sim$50 GHz peak is dominated by the thermal free-free emission and the nonthermal synchrotron radio emission associated with massive star formation.

Another spectrally peaked structure for the radio continuum ($\sim$10--90 GHz) emission is anomalous microwave emission (AME;  \citealt{Dickinson2018}), which seems ubiquitous in the Galaxy (e.g., \citealt{Kogut1996}; \citealt{Davies2006}; \citealt{PlanckCollaboration2016}) but has surprisingly few detections in star-forming regions outside of the Galaxy (e.g., \citealt{Murphy2010, Murphy2018}). Anomalous microwave emission is found spatially correlated with FIR thermal dust emission. The rest frame $\sim$50 GHz emission of J1146+4037 comes from the entire quasar and cannot be resolved. In addition, it is too bright for an AME nature. For example, \citet{Hensley2016} present a linear correlation between the 30 GHz AME intensity and the dust radiance by investigating the diffuse Galactic AME observed by Planck and the FIR dust emission. With their conversion factor and the 3$\sigma$ flux density upper limit  from the JCMT 850 $\mu$m observations, we calculate a 30 GHz AME flux density upper limit of 0.00045 mJy. This is far smaller than the value ($\sim$1 mJy in the rest frame) we detect.

\subsubsection{Blazar flaring?}
\label{bf}
Considering the blazar nature of J1146+4037, the two spectral peaks may both be due to blazar flaring. Studies demonstrate that blazars with ongoing flaring can have peaked radio spectra at high frequencies (e.g.,  \citealt{ODea1986}; \citealt{Tinti2005}). 
At observed frame 5 GHz ($\sim$30 GHz in the rest frame), \citet{Frey2010} observed this target  on 2008 October 22 at a resolution of $5.1\times1.4$ mas with the EVN, and find a compact morphology with flux density of $8.6\pm0.4$ mJy. Our $\sim$$0\farcs3$ VLA observations reveal a flux density of $9.57\pm0.07$ mJy beam$^{-1}$ at observed frame 5 GHz. Considering the calibration error, these two measurements are consistent, indicating that almost all the radio emission is from the EVN-detected compact region, and that this quasar is in a rather stable state  during about $\sim$10 years. 
Long-term multiwavelength monitoring toward J1146+4037 is necessary to check if these two spectral peaks showing in Fig. \ref{J1146modelfit}  are caused by intermittent AGN activity.

\subsubsection{Two peaks from multiple components?}
In addition to blazar flaring, we now discuss a multicomponent origin for the two spectral peaks. For example, radio sources with core-jet morphology may produce such a two-peaked spectrum.
\citet{Frey2011} proposed a likely compact nature toward this target by milliarcsecond-resolution EVN observations. As discussed in Sect. \ref{bf}, comparing the archival milliarcsecond-resolution observations with our sub-arcsecond resolution observations, the flux loss may be negligible. This suggests the absence of  strong radio emission outside the EVN detected compact region toward this target, while we can still see irregular flux density contours on the 5 GHz image in the Fig. 2 of \citet{Frey2011}, which may indicate the existence of a weak jet. This is consistent with its blazar nature (i.e., its small viewing angle of 3$\degr$; \citealt{Ghisellini2014}), which allows the relativistic jet to be observed (viewing angle\,$\ne$\,0$\degr$) but hard to detect with 5 GHz EVN resolution. Higher frequency (and therefore higher angular resolution) EVN or VLBA observations are needed to see if the central radio emission can be resolved into multiple components.

\section{Summary} 
\label{sum}
We report on VLA S-, C- and X-band continuum observations toward seven radio-loud quasars at $z>5$  that we selected from our previous GMRT project, which presented the 323 MHz radio continuum of 13 radio-loud quasars at $z>5$ \citep{Shao2020}, to investigate if there is any spectral turnover at  frequencies $>1$ GHz in the observed frame.
We also present  uGMRT band-2, band-3, and band-4 radio continuum observations toward eight radio-loud quasars at $z>5$, also selected  from our previous GMRT project, to sample their low-frequency synchrotron emission. 
Below are our main results:\begin{itemize}
\item All seven VLA targets are detected as point sources with  S/N$_{\rm peak}$ $\sim$5$-$1000$\sigma$.
For the eight uGMRT targets, all band-2 observations are influenced by very damaging RFI. The remaining band-3 and band-4 observations are successful, and all sources are detected as point sources with S/N$_{\rm peak}$ $\sim$8--106$\sigma$. 

\item Combined with previous radio continuum observations from the literature, nine targets show obvious evidence of spectral turnover and another one has a  tentative and less constrained spectral turnover. The turnover frequencies are $\sim$1--50 GHz for our sample in the rest frame. This makes our targets  GPS or HFP candidates. All targets have enough data on both sides of the spectral peaks, except for J2228+0110, which is less constrained with a low-frequency upper limit. J1146+4037 shows a complex spectrum with two spectral peaks.  

\item The turnovers observed in the radio spectra of our targets may be an artifact of source variability due to the intrinsically variable nature of quasars. 
We assume that the observed spectral turnovers are genuine and apply a series of absorption models associated with SSA and FFA to the entire radio spectra toward the nine well-constrained targets. All the targets can be accurately fit with the FFA in an external inhomogeneous medium case. 
This may remove an SSA case and indicate an FFA origin of the spectral turnover of our sample. The mismatch of the magnetic field strength from equipartition and the SSA assumption toward J0131--0321 and J1146+4037 are also in contradiction with an SSA origin. We derive the mean densities of the absorbing clumpy ISM, under the assumption of log-normal distributions of warm gas, to be thousands to tens of thousands of cm$^{-3}$ for J0131--0321 and J1146+4037.

\item As for the complex radio spectrum of J1146+4037 with two spectral peaks: We argue that the two spectral peaks may be from multiple components (e.g., core-jet) and caused by FFA from the high-density medium in the nuclear region. However, we cannot rule out the possibility of quasar variability. High-frequency milliarcsecond observations with high sensitivity and multi-epoch monitoring of the entire radio spectrum toward this target are needed to check these scenarios.

\item From our radio spectral modeling, we calculate the radio loudness $R_{2500\rm\, \AA}$ toward our sample, which ranges from 12$^{+1}_{-1}$ to 674$^{+61}_{-51}$.  This is the first precise radio loudness with S/N $\sim$6--16$\sigma$ released for such a large sample size for radio-loud quasars at $z>5$.

\end{itemize}

The radio spectral turnover at rest frame $\sim$1--50 GHz for our $z>5$ radio-loud quasar sample may be caused by FFA in an external inhomogeneous medium; however, the origin of variability is also possible.

\begin{acknowledgements}
We thank the VLA and GMRT staff who helped to observe our projects. The National Radio Astronomy Observatory is a facility of the National Science Foundation operated under cooperative agreement by Associated Universities, Inc. The GMRT is run by the National Centre for Radio Astrophysics of the Tata Institute of Fundamental Research. R.W. acknowledges supports from the National Science Foundation of China (NSFC)  grants No. 11991052, 11721303, and 11533001. D.R. acknowledges support from the National Science Foundation under grant number AST-1614213 and AST-1910107 and from the Alexander von Humboldt Foundation through a Humboldt Research Fellowship for Experienced Researchers. 
\end{acknowledgements}

\clearpage
\bibliographystyle{aa} 
\bibliography{my}

\begin{thebibliography}{114}
\expandafter\ifx\csname natexlab\endcsname\relax\def\natexlab#1{#1}\fi

\bibitem[{{An} \& {Baan}(2012)}]{An2012}
{An}, T. \& {Baan}, W.~A. 2012, \apj, 760, 77

\bibitem[{{Anderson} {et~al.}(2001){Anderson}, {Fan}, {Richards}, {Schneider},
  {Strauss}, {Vanden Berk}, {Gunn}, {Knapp}, {Schlegel}, {Voges}, {Yanny},
  {Bahcall}, {Bernardi}, {Brinkmann}, {Brunner}, {Csab{\'a}i}, {Doi},
  {Fukugita}, {Hennessy}, {Ivezi{\'c}}, {Kunszt}, {Lamb}, {Loveday}, {Lupton},
  {McKay}, {Munn}, {Nichol}, {Szokoly}, \& {York}}]{Anderson2001}
{Anderson}, S.~F., {Fan}, X., {Richards}, G.~T., {et~al.} 2001, \aj, 122, 503

\bibitem[{{Ba{\~n}ados} {et~al.}(2018){Ba{\~n}ados}, {Carilli}, {Walter},
  {Momjian}, {Decarli}, {Farina}, {Mazzucchelli}, \& {Venemans}}]{Banados2018}
{Ba{\~n}ados}, E., {Carilli}, C., {Walter}, F., {et~al.} 2018, \apjl, 861, L14

\bibitem[{{Ba{\~n}ados} {et~al.}(2021){Ba{\~n}ados}, {Mazzucchelli}, {Momjian},
  {Eilers}, {Wang}, {Schindler}, {Connor}, {Andika}, {Barth}, {Carilli},
  {Davies}, {Decarli}, {Fan}, {Farina}, {Hennawi}, {Pensabene}, {Stern},
  {Venemans}, {Wenzl}, \& {Yang}}]{Banados2021}
{Ba{\~n}ados}, E., {Mazzucchelli}, C., {Momjian}, E., {et~al.} 2021, \apj, 909,
  80

\bibitem[{{Ba{\~n}ados} {et~al.}(2015){Ba{\~n}ados}, {Venemans}, {Morganson},
  {Hodge}, {Decarli}, {Walter}, {Stern}, {Schlafly}, {Farina}, {Greiner},
  {Chambers}, {Fan}, {Rix}, {Burgett}, {Draper}, {Flewelling}, {Kaiser},
  {Metcalfe}, {Morgan}, {Tonry}, \& {Wainscoat}}]{Banados2015}
{Ba{\~n}ados}, E., {Venemans}, B.~P., {Morganson}, E., {et~al.} 2015, \apj,
  804, 118

\bibitem[{{Beelen} {et~al.}(2006){Beelen}, {Cox}, {Benford}, {Dowell},
  {Kov{\'a}cs}, {Bertoldi}, {Omont}, \& {Carilli}}]{Beelen2006}
{Beelen}, A., {Cox}, P., {Benford}, D.~J., {et~al.} 2006, \apj, 642, 694

\bibitem[{{Begelman}(1999)}]{Begelman1999}
{Begelman}, M.~C. 1999, in The Most Distant Radio Galaxies, ed. H.~J.~A.
  {R{\"o}ttgering}, P.~N. {Best}, \& M.~D. {Lehnert}, 173

\bibitem[{{Belladitta} {et~al.}(2020){Belladitta}, {Moretti}, {Caccianiga},
  {Spingola}, {Severgnini}, {Della Ceca}, {Ghisellini}, {Dallacasa},
  {Sbarrato}, {Cicone}, {Cassar{\`a}}, \& {Pedani}}]{Belladitta2020}
{Belladitta}, S., {Moretti}, A., {Caccianiga}, A., {et~al.} 2020, \aap, 635, L7

\bibitem[{{Bicknell} {et~al.}(1997){Bicknell}, {Dopita}, \&
  {O'Dea}}]{Bicknell1997}
{Bicknell}, G.~V., {Dopita}, M.~A., \& {O'Dea}, C.~P.~O. 1997, \apj, 485, 112

\bibitem[{{Bicknell} {et~al.}(2018){Bicknell}, {Mukherjee}, {Wagner},
  {Sutherland}, \& {Nesvadba}}]{Bicknell2018}
{Bicknell}, G.~V., {Mukherjee}, D., {Wagner}, A.~Y., {Sutherland}, R.~S., \&
  {Nesvadba}, N. P.~H. 2018, \mnras, 475, 3493

\bibitem[{{Callingham} {et~al.}(2015){Callingham}, {Gaensler}, {Ekers},
  {Tingay}, {Wayth}, {Morgan}, {Bernardi}, {Bell}, {Bhat}, {Bowman}, {Briggs},
  {Cappallo}, {Deshpande}, {Ewall-Wice}, {Feng}, {Greenhill}, {Hazelton},
  {Hindson}, {Hurley-Walker}, {Jacobs}, {Johnston-Hollitt}, {Kaplan},
  {Kudrayvtseva}, {Lenc}, {Lonsdale}, {McKinley}, {McWhirter}, {Mitchell},
  {Morales}, {Morgan}, {Oberoi}, {Offringa}, {Ord}, {Pindor}, {Prabu},
  {Procopio}, {Riding}, {Srivani}, {Subrahmanyan}, {Udaya Shankar}, {Webster},
  {Williams}, \& {Williams}}]{Callingham2015}
{Callingham}, J.~R., {Gaensler}, B.~M., {Ekers}, R.~D., {et~al.} 2015, \apj,
  809, 168

\bibitem[{{Cao} {et~al.}(2014){Cao}, {Frey}, {Gurvits}, {Yang}, {Hong},
  {Paragi}, {Deller}, \& {Ivezi{\'c}}}]{Cao2014}
{Cao}, H.-M., {Frey}, S., {Gurvits}, L.~I., {et~al.} 2014, \aap, 563, A111

\bibitem[{{Carilli} \& {Barthel}(1996)}]{Carilli1996}
{Carilli}, C.~L. \& {Barthel}, P.~D. 1996, \aapr, 7, 1

\bibitem[{{Carilli} {et~al.}(2007){Carilli}, {Wang}, {van Hoven},
  {Dwarakanath}, {Chengalur}, \& {Wyithe}}]{Carilli2007}
{Carilli}, C.~L., {Wang}, R., {van Hoven}, M.~B., {et~al.} 2007, \aj, 133, 2841

\bibitem[{{Carvalho}(1998)}]{Carvalho1998}
{Carvalho}, J.~C. 1998, \aap, 329, 845

\bibitem[{{Chapin} {et~al.}(2013){Chapin}, {Berry}, {Gibb}, {Jenness}, {Scott},
  {Tilanus}, {Economou}, \& {Holland}}]{Chapin2013}
{Chapin}, E.~L., {Berry}, D.~S., {Gibb}, A.~G., {et~al.} 2013, \mnras, 430,
  2545

\bibitem[{{Condon}(1992)}]{Condon1992}
{Condon}, J.~J. 1992, \araa, 30, 575

\bibitem[{{Coppejans} {et~al.}(2015){Coppejans}, {Cseh}, {Williams}, {van
  Velzen}, \& {Falcke}}]{Coppejans2015}
{Coppejans}, R., {Cseh}, D., {Williams}, W.~L., {van Velzen}, S., \& {Falcke},
  H. 2015, \mnras, 450, 1477

\bibitem[{{Coppejans} {et~al.}(2017){Coppejans}, {van Velzen}, {Intema},
  {M{\"u}ller}, {Frey}, {Coppejans}, {Cseh}, {Williams}, {Falcke},
  {K{\"o}rding}, {Orr{\'u}}, {Paragi}, \& {Gab{\'a}nyi}}]{Coppejans2017}
{Coppejans}, R., {van Velzen}, S., {Intema}, H.~T., {et~al.} 2017, \mnras, 467,
  2039

\bibitem[{{Dallacasa} {et~al.}(2000){Dallacasa}, {Stanghellini}, {Centonza}, \&
  {Fanti}}]{Dallacasa2000}
{Dallacasa}, D., {Stanghellini}, C., {Centonza}, M., \& {Fanti}, R. 2000, \aap,
  363, 887

\bibitem[{{Dasyra} \& {Combes}(2012)}]{Dasyra2012}
{Dasyra}, K.~M. \& {Combes}, F. 2012, \aap, 541, L7

\bibitem[{{Davies} {et~al.}(2006){Davies}, {Dickinson}, {Banday}, {Jaffe},
  {G{\'o}rski}, \& {Davis}}]{Davies2006}
{Davies}, R.~D., {Dickinson}, C., {Banday}, A.~J., {et~al.} 2006, \mnras, 370,
  1125

\bibitem[{{de Vries} {et~al.}(2009){de Vries}, {Snellen}, {Schilizzi}, \&
  {Mack}}]{deVries2009}
{de Vries}, N., {Snellen}, I.~A.~G., {Schilizzi}, R.~T., \& {Mack}, K.~H. 2009,
  Astronomische Nachrichten, 330, 214

\bibitem[{{De Young}(1991)}]{DeYoung1991}
{De Young}, D.~S. 1991, \apj, 371, 69

\bibitem[{{De Young}(1993)}]{DeYoung1993}
{De Young}, D.~S. 1993, \apj, 402, 95

\bibitem[{{Dempsey} {et~al.}(2013){Dempsey}, {Friberg}, {Jenness}, {Tilanus},
  {Thomas}, {Holland}, {Bintley}, {Berry}, {Chapin}, {Chrysostomou}, {Davis},
  {Gibb}, {Parsons}, \& {Robson}}]{Dempsey2013}
{Dempsey}, J.~T., {Friberg}, P., {Jenness}, T., {et~al.} 2013, \mnras, 430,
  2534

\bibitem[{{Dickinson} {et~al.}(2018){Dickinson}, {Ali-Ha{\"\i}moud}, {Barr},
  {Battistelli}, {Bell}, {Bernstein}, {Casassus}, {Cleary}, {Draine},
  {G{\'e}nova-Santos}, {Harper}, {Hensley}, {Hill-Valler}, {Hoang}, {Israel},
  {Jew}, {Lazarian}, {Leahy}, {Leech}, {L{\'o}pez-Caraballo}, {McDonald},
  {Murphy}, {Onaka}, {Paladini}, {Peel}, {Perrott}, {Poidevin}, {Readhead},
  {Rubi{\~n}o-Mart{\'\i}n}, {Taylor}, {Tibbs}, {Todorovi{\'c}}, \&
  {Vidal}}]{Dickinson2018}
{Dickinson}, C., {Ali-Ha{\"\i}moud}, Y., {Barr}, A., {et~al.} 2018, \nar, 80, 1

\bibitem[{{Evans} {et~al.}(1999){Evans}, {Kim}, {Mazzarella}, {Scoville}, \&
  {Sanders}}]{Evans1999}
{Evans}, A.~S., {Kim}, D.~C., {Mazzarella}, J.~M., {Scoville}, N.~Z., \&
  {Sanders}, D.~B. 1999, \apjl, 521, L107

\bibitem[{{Fan} {et~al.}(2001){Fan}, {Narayanan}, {Lupton}, {Strauss}, {Knapp},
  {Becker}, {White}, {Pentericci}, {Leggett}, {Haiman}, {Gunn}, {Ivezi{\'c}},
  {Schneider}, {Anderson}, {Brinkmann}, {Bahcall}, {Connolly}, {Csabai}, {Doi},
  {Fukugita}, {Geballe}, {Grebel}, {Harbeck}, {Hennessy}, {Lamb}, {Miknaitis},
  {Munn}, {Nichol}, {Okamura}, {Pier}, {Prada}, {Richards}, {Szalay}, \&
  {York}}]{Fan2001}
{Fan}, X., {Narayanan}, V.~K., {Lupton}, R.~H., {et~al.} 2001, \aj, 122, 2833

\bibitem[{{Fanti} {et~al.}(1990){Fanti}, {Fanti}, {Schilizzi}, {Spencer}, {Nan
  Rendong}, {Parma}, {van Breugel}, \& {Venturi}}]{Fanti1990}
{Fanti}, R., {Fanti}, C., {Schilizzi}, R.~T., {et~al.} 1990, \aap, 231, 333

\bibitem[{{Flesch}(2021)}]{Flesch2021}
{Flesch}, E.~W. 2021, arXiv e-prints, arXiv:2105.12985

\bibitem[{{Foreman-Mackey} {et~al.}(2013){Foreman-Mackey}, {Hogg}, {Lang}, \&
  {Goodman}}]{emcee2013}
{Foreman-Mackey}, D., {Hogg}, D.~W., {Lang}, D., \& {Goodman}, J. 2013, \pasp,
  125, 306

\bibitem[{{F{\"o}rster Schreiber} {et~al.}(2009){F{\"o}rster Schreiber},
  {Genzel}, {Bouch{\'e}}, {Cresci}, {Davies}, {Buschkamp}, {Shapiro},
  {Tacconi}, {Hicks}, {Genel}, {Shapley}, {Erb}, {Steidel}, {Lutz},
  {Eisenhauer}, {Gillessen}, {Sternberg}, {Renzini}, {Cimatti}, {Daddi},
  {Kurk}, {Lilly}, {Kong}, {Lehnert}, {Nesvadba}, {Verma}, {McCracken},
  {Arimoto}, {Mignoli}, \& {Onodera}}]{ForsterSchreiber2009}
{F{\"o}rster Schreiber}, N.~M., {Genzel}, R., {Bouch{\'e}}, N., {et~al.} 2009,
  \apj, 706, 1364

\bibitem[{{Frey} {et~al.}(2008){Frey}, {Gurvits}, {Paragi}, \&
  {{\'E}.~Gab{\'a}nyi}}]{Frey2008}
{Frey}, S., {Gurvits}, L.~I., {Paragi}, Z., \& {{\'E}.~Gab{\'a}nyi}, K. 2008,
  \aap, 484, L39

\bibitem[{{Frey} {et~al.}(2003){Frey}, {Mosoni}, {Paragi}, \&
  {Gurvits}}]{Frey2003}
{Frey}, S., {Mosoni}, L., {Paragi}, Z., \& {Gurvits}, L.~I. 2003, \mnras, 343,
  L20

\bibitem[{{Frey} {et~al.}(2015){Frey}, {Paragi}, {Fogasy}, \&
  {Gurvits}}]{Frey2015}
{Frey}, S., {Paragi}, Z., {Fogasy}, J.~O., \& {Gurvits}, L.~I. 2015, \mnras,
  446, 2921

\bibitem[{{Frey} {et~al.}(2010){Frey}, {Paragi}, {Gurvits}, {Cseh}, \&
  {Gab{\'a}nyi}}]{Frey2010}
{Frey}, S., {Paragi}, Z., {Gurvits}, L.~I., {Cseh}, D., \& {Gab{\'a}nyi},
  K.~{\'E}. 2010, \aap, 524, A83

\bibitem[{{Frey} {et~al.}(2011){Frey}, {Paragi}, {Gurvits}, {Gab{\'a}nyi}, \&
  {Cseh}}]{Frey2011}
{Frey}, S., {Paragi}, Z., {Gurvits}, L.~I., {Gab{\'a}nyi}, K.~{\'E}., \&
  {Cseh}, D. 2011, \aap, 531, L5

\bibitem[{{Frey} {et~al.}(2005){Frey}, {Paragi}, {Mosoni}, \&
  {Gurvits}}]{Frey2005}
{Frey}, S., {Paragi}, Z., {Mosoni}, L., \& {Gurvits}, L.~I. 2005, \aap, 436,
  L13

\bibitem[{{Gab{\'a}nyi} {et~al.}(2015){Gab{\'a}nyi}, {Cseh}, {Frey}, {Paragi},
  {Gurvits}, {An}, \& {Zhang}}]{Gabanyi2015}
{Gab{\'a}nyi}, K.~{\'E}., {Cseh}, D., {Frey}, S., {et~al.} 2015, \mnras, 450,
  L57

\bibitem[{{Ghisellini} {et~al.}(2014){Ghisellini}, {Sbarrato}, {Tagliaferri},
  {Foschini}, {Tavecchio}, {Ghirlanda}, {Braito}, \&
  {Gehrels}}]{Ghisellini2014}
{Ghisellini}, G., {Sbarrato}, T., {Tagliaferri}, G., {et~al.} 2014, \mnras,
  440, L111

\bibitem[{{Gopal-Krishna} {et~al.}(1983){Gopal-Krishna}, {Patnaik}, \&
  {Steppe}}]{Gopal-Krishna1983}
{Gopal-Krishna}, {Patnaik}, A.~R., \& {Steppe}, H. 1983, \aap, 123, 107

\bibitem[{{Gupta} {et~al.}(2021){Gupta}, {Shukla}, {Srianand}, {Krogager},
  {Noterdaeme}, {Baker}, {Combes}, {Fynbo}, {Momjian}, {Hilton}, {Hussain},
  {Moodley}, {Petitjean}, {Chen}, {Deka}, {Dutta}, {Jose}, {Jozsa}, {Kaski},
  {Klockner}, {Knowles}, {Sikhosana}, \& {Wagenveld}}]{Gupta2021}
{Gupta}, N., {Shukla}, G., {Srianand}, R., {et~al.} 2021, arXiv e-prints,
  arXiv:2107.09705

\bibitem[{{Hajela} {et~al.}(2019){Hajela}, {Mooley}, {Intema}, \&
  {Frail}}]{Hajela2019}
{Hajela}, A., {Mooley}, K.~P., {Intema}, H.~T., \& {Frail}, D.~A. 2019, \mnras,
  490, 4898

\bibitem[{{Helfand} {et~al.}(2015){Helfand}, {White}, \&
  {Becker}}]{Helfand2015}
{Helfand}, D.~J., {White}, R.~L., \& {Becker}, R.~H. 2015, \apj, 801, 26

\bibitem[{{Hensley} {et~al.}(2016){Hensley}, {Draine}, \&
  {Meisner}}]{Hensley2016}
{Hensley}, B.~S., {Draine}, B.~T., \& {Meisner}, A.~M. 2016, \apj, 827, 45

\bibitem[{{Hewett} \& {Wild}(2010)}]{Hewett2010}
{Hewett}, P.~C. \& {Wild}, V. 2010, \mnras, 405, 2302

\bibitem[{{Hodge} {et~al.}(2011){Hodge}, {Becker}, {White}, {Richards}, \&
  {Zeimann}}]{Hodge2011}
{Hodge}, J.~A., {Becker}, R.~H., {White}, R.~L., {Richards}, G.~T., \&
  {Zeimann}, G.~R. 2011, \aj, 142, 3

\bibitem[{{Ighina} {et~al.}(2021){Ighina}, {Belladitta}, {Caccianiga},
  {Broderick}, {Drouart}, {Moretti}, \& {Seymour}}]{Ighina2021}
{Ighina}, L., {Belladitta}, S., {Caccianiga}, A., {et~al.} 2021, \aap, 647, L11

\bibitem[{{Intema} {et~al.}(2017){Intema}, {Jagannathan}, {Mooley}, \&
  {Frail}}]{Intema2017}
{Intema}, H.~T., {Jagannathan}, P., {Mooley}, K.~P., \& {Frail}, D.~A. 2017,
  \aap, 598, A78

\bibitem[{{Jauncey} {et~al.}(2020){Jauncey}, {Koay}, {Bignall}, {Macquart},
  {Pursimo}, {Giroletti}, {Hovatta}, {Kiehlmann}, {Rickett}, {Readhead},
  {Max-Moerbeck}, {Vedantham}, {Reynolds}, {Lovell}, {Ojha}, \&
  {Kedziora-Chudczer}}]{Jauncey2020}
{Jauncey}, D.~L., {Koay}, J.~Y., {Bignall}, H., {et~al.} 2020, Advances in
  Space Research, 65, 756

\bibitem[{{Jeyakumar}(2016)}]{Jeyakumar2016}
{Jeyakumar}, S. 2016, \mnras, 458, 3786

\bibitem[{{Junor} {et~al.}(1999){Junor}, {Salter}, {Saikia}, {Mantovani}, \&
  {Peck}}]{Junor1999}
{Junor}, W., {Salter}, C.~J., {Saikia}, D.~J., {Mantovani}, F., \& {Peck},
  A.~B. 1999, \mnras, 308, 955

\bibitem[{{Kale} \& {Ishwara-Chandra}(2020)}]{Kale2020}
{Kale}, R. \& {Ishwara-Chandra}, C.~H. 2020, {CAPTURE: Interferometric pipeline
  for image creation from GMRT data}

\bibitem[{{Kellermann}(1966)}]{Kellermann1966}
{Kellermann}, K.~I. 1966, \apj, 146, 621

\bibitem[{{Kellermann} \& {Pauliny-Toth}(1981)}]{Kellermann1981}
{Kellermann}, K.~I. \& {Pauliny-Toth}, I.~I.~K. 1981, \araa, 19, 373

\bibitem[{{Kellermann} {et~al.}(1989){Kellermann}, {Sramek}, {Schmidt},
  {Shaffer}, \& {Green}}]{Kellermann1989}
{Kellermann}, K.~I., {Sramek}, R., {Schmidt}, M., {Shaffer}, D.~B., \& {Green},
  R. 1989, \aj, 98, 1195

\bibitem[{{Khorunzhev} {et~al.}(2021){Khorunzhev}, {Meshcheryakov}, {Medvedev},
  {Borisov}, {Burenin}, {Krivonos}, {Uklein}, {Shablovinskaya}, {Afanasiev},
  {Dodonov}, {Sunyaev}, {Sazonov}, \& {Gilfanov}}]{Khorunzhev2021}
{Khorunzhev}, G.~A., {Meshcheryakov}, A.~V., {Medvedev}, P.~S., {et~al.} 2021,
  Astronomy Letters, 47, 123

\bibitem[{{Kogut} {et~al.}(1996){Kogut}, {Banday}, {Bennett}, {Gorski},
  {Hinshaw}, \& {Reach}}]{Kogut1996}
{Kogut}, A., {Banday}, A.~J., {Bennett}, C.~L., {et~al.} 1996, \apj, 460, 1

\bibitem[{{Kovalev}(2005)}]{Kovalev2005}
{Kovalev}, Y.~Y. 2005, Baltic Astronomy, 14, 413

\bibitem[{{Kovalev} {et~al.}(2002){Kovalev}, {Kovalev}, {Nizhelsky}, \&
  {Bogdantsov}}]{Kovalev2002}
{Kovalev}, Y.~Y., {Kovalev}, Y.~A., {Nizhelsky}, N.~A., \& {Bogdantsov}, A.~B.
  2002, \pasa, 19, 83

\bibitem[{{Leipski} {et~al.}(2014){Leipski}, {Meisenheimer}, {Walter}, {Klaas},
  {Dannerbauer}, {De Rosa}, {Fan}, {Haas}, {Krause}, \& {Rix}}]{Leipski2014}
{Leipski}, C., {Meisenheimer}, K., {Walter}, F., {et~al.} 2014, \apj, 785, 154

\bibitem[{{Liu} {et~al.}(2021){Liu}, {Wang}, {Momjian}, {Ba{\~n}ados},
  {Zeimann}, {Willott}, {Matsuoka}, {Omont}, {Shao}, {Li}, \& {Li}}]{Liu2021}
{Liu}, Y., {Wang}, R., {Momjian}, E., {et~al.} 2021, \apj, 908, 124

\bibitem[{{Maciel} \& {Alexander}(2014)}]{Maciel2014}
{Maciel}, T. \& {Alexander}, P. 2014, \mnras, 442, 3469

\bibitem[{{McGreer} {et~al.}(2006){McGreer}, {Becker}, {Helfand}, \&
  {White}}]{McGreer2006}
{McGreer}, I.~D., {Becker}, R.~H., {Helfand}, D.~J., \& {White}, R.~L. 2006,
  \apj, 652, 157

\bibitem[{{McGreer} {et~al.}(2009){McGreer}, {Helfand}, \&
  {White}}]{McGreer2009}
{McGreer}, I.~D., {Helfand}, D.~J., \& {White}, R.~L. 2009, \aj, 138, 1925

\bibitem[{{Mezger} \& {Henderson}(1967)}]{Mezger1967}
{Mezger}, P.~G. \& {Henderson}, A.~P. 1967, \apj, 147, 471

\bibitem[{{Mhaskey} {et~al.}(2019){Mhaskey}, {Gopal-Krishna}, {Paul},
  {Dabhade}, {Salunkhe}, {Bhagat}, \& {Bendre}}]{Mhaskey2019}
{Mhaskey}, M., {Gopal-Krishna}, {Paul}, S., {et~al.} 2019, \mnras, 489, 3506

\bibitem[{{Momjian} {et~al.}(2021){Momjian}, {Ba{\~n}ados}, {Carilli},
  {Walter}, \& {Mazzucchelli}}]{Momjian2021}
{Momjian}, E., {Ba{\~n}ados}, E., {Carilli}, C.~L., {Walter}, F., \&
  {Mazzucchelli}, C. 2021, \aj, 161, 207

\bibitem[{{Momjian} {et~al.}(2018){Momjian}, {Carilli}, {Ba{\~n}ados},
  {Walter}, \& {Venemans}}]{Momjian2018}
{Momjian}, E., {Carilli}, C.~L., {Ba{\~n}ados}, E., {Walter}, F., \&
  {Venemans}, B.~P. 2018, \apj, 861, 86

\bibitem[{{Momjian} {et~al.}(2008){Momjian}, {Carilli}, \&
  {McGreer}}]{Momjian2008}
{Momjian}, E., {Carilli}, C.~L., \& {McGreer}, I.~D. 2008, \aj, 136, 344

\bibitem[{{Momjian} {et~al.}(2003){Momjian}, {Petric}, \&
  {Carilli}}]{Momjian2003}
{Momjian}, E., {Petric}, A.~O., \& {Carilli}, C.~L. 2003, in Bulletin of the
  American Astronomical Society, Vol.~35, American Astronomical Society Meeting
  Abstracts, 1327

\bibitem[{{Morganti} {et~al.}(2021){Morganti}, {Oosterloo}, \&
  {Tadhunter}}]{Morganti2021}
{Morganti}, R., {Oosterloo}, T., \& {Tadhunter}, C.~N. 2021, IAU Symposium,
  359, 243

\bibitem[{{Murphy} {et~al.}(2012){Murphy}, {Bremseth}, {Mason}, {Condon},
  {Schinnerer}, {Aniano}, {Armus}, {Helou}, {Turner}, \&
  {Jarrett}}]{Murphy2012}
{Murphy}, E.~J., {Bremseth}, J., {Mason}, B.~S., {et~al.} 2012, \apj, 761, 97

\bibitem[{{Murphy} {et~al.}(2010{\natexlab{a}}){Murphy}, {Helou}, {Condon},
  {Schinnerer}, {Turner}, {Beck}, {Mason}, {Chary}, \& {Armus}}]{Murphy2010}
{Murphy}, E.~J., {Helou}, G., {Condon}, J.~J., {et~al.} 2010{\natexlab{a}},
  \apjl, 709, L108

\bibitem[{{Murphy} {et~al.}(2018){Murphy}, {Linden}, {Dong}, {Hensley},
  {Momjian}, {Helou}, \& {Evans}}]{Murphy2018}
{Murphy}, E.~J., {Linden}, S.~T., {Dong}, D., {et~al.} 2018, \apj, 862, 20

\bibitem[{{Murphy} {et~al.}(2010{\natexlab{b}}){Murphy}, {Cohen}, {Ekers},
  {Green}, {Wark}, \& {Moss}}]{MurphyTara2010}
{Murphy}, T., {Cohen}, M., {Ekers}, R.~D., {et~al.} 2010{\natexlab{b}}, \mnras,
  405, 1560

\bibitem[{{O'Dea}(1998)}]{ODea1998}
{O'Dea}, C.~P. 1998, \pasp, 110, 493

\bibitem[{{O'Dea} {et~al.}(1983){O'Dea}, {Dent}, {Balonek}, \&
  {Kapitzky}}]{ODea1983}
{O'Dea}, C.~P., {Dent}, W.~A., {Balonek}, T.~J., \& {Kapitzky}, J.~E. 1983,
  \aj, 88, 1616

\bibitem[{{O'Dea} {et~al.}(1986){O'Dea}, {Dent}, {Kinzel}, \&
  {Balonek}}]{ODea1986}
{O'Dea}, C.~P., {Dent}, W.~A., {Kinzel}, W.~M., \& {Balonek}, T.~J. 1986, \aj,
  92, 1262

\bibitem[{{O'Dea} \& {Saikia}(2021)}]{ODea2021}
{O'Dea}, C.~P. \& {Saikia}, D.~J. 2021, \aapr, 29, 3

\bibitem[{{Orienti} \& {Dallacasa}(2008)}]{Orienti2008}
{Orienti}, M. \& {Dallacasa}, D. 2008, \aap, 487, 885

\bibitem[{{Orienti} \& {Dallacasa}(2012)}]{Orienti2012}
{Orienti}, M. \& {Dallacasa}, D. 2012, \mnras, 424, 532

\bibitem[{{Orienti} {et~al.}(2007){Orienti}, {Dallacasa}, \&
  {Stanghellini}}]{Orienti2007}
{Orienti}, M., {Dallacasa}, D., \& {Stanghellini}, C. 2007, \aap, 461, 923

\bibitem[{{Ostorero} {et~al.}(2017){Ostorero}, {Morganti}, {Diaferio},
  {Siemiginowska}, {Stawarz}, {Moderski}, \& {Labiano}}]{Ostorero2017}
{Ostorero}, L., {Morganti}, R., {Diaferio}, A., {et~al.} 2017, \apj, 849, 34

\bibitem[{{Pacholczyk}(1970)}]{Pacholczyk1970}
{Pacholczyk}, A.~G. 1970, {Radio astrophysics. Nonthermal processes in galactic
  and extragalactic sources}

\bibitem[{{Peck} {et~al.}(1999){Peck}, {Taylor}, \& {Conway}}]{Peck1999}
{Peck}, A.~B., {Taylor}, G.~B., \& {Conway}, J.~E. 1999, \apj, 521, 103

\bibitem[{{Petric} {et~al.}(2003){Petric}, {Carilli}, {Bertoldi}, {Fan}, {Cox},
  {Strauss}, {Omont}, \& {Schneider}}]{Petric2003}
{Petric}, A.~O., {Carilli}, C.~L., {Bertoldi}, F., {et~al.} 2003, \aj, 126, 15

\bibitem[{{Planck Collaboration} {et~al.}(2016){Planck Collaboration}, {Ade},
  {Aghanim}, {Arnaud}, {Ashdown}, {Aumont}, {Baccigalupi}, {Banday},
  {Barreiro}, {Bartlett}, {Bartolo}, {Battaner}, {Battye}, {Benabed},
  {Beno{\^\i}t}, {Benoit-L{\'e}vy}, {Bernard}, {Bersanelli}, {Bielewicz},
  {Bock}, {Bonaldi}, {Bonavera}, {Bond}, {Borrill}, {Bouchet}, {Boulanger},
  {Bucher}, {Burigana}, {Butler}, {Calabrese}, {Cardoso}, {Catalano},
  {Challinor}, {Chamballu}, {Chary}, {Chiang}, {Chluba}, {Christensen},
  {Church}, {Clements}, {Colombi}, {Colombo}, {Combet}, {Coulais}, {Crill},
  {Curto}, {Cuttaia}, {Danese}, {Davies}, {Davis}, {de Bernardis}, {de Rosa},
  {de Zotti}, {Delabrouille}, {D{\'e}sert}, {Di Valentino}, {Dickinson},
  {Diego}, {Dolag}, {Dole}, {Donzelli}, {Dor{\'e}}, {Douspis}, {Ducout},
  {Dunkley}, {Dupac}, {Efstathiou}, {Elsner}, {En{\ss}lin}, {Eriksen},
  {Farhang}, {Fergusson}, {Finelli}, {Forni}, {Frailis}, {Fraisse},
  {Franceschi}, {Frejsel}, {Galeotta}, {Galli}, {Ganga}, {Gauthier}, {Gerbino},
  {Ghosh}, {Giard}, {Giraud-H{\'e}raud}, {Giusarma}, {Gjerl{\o}w},
  {Gonz{\'a}lez-Nuevo}, {G{\'o}rski}, {Gratton}, {Gregorio}, {Gruppuso},
  {Gudmundsson}, {Hamann}, {Hansen}, {Hanson}, {Harrison}, {Helou},
  {Henrot-Versill{\'e}}, {Hern{\'a}ndez-Monteagudo}, {Herranz}, {Hildebrand t},
  {Hivon}, {Hobson}, {Holmes}, {Hornstrup}, {Hovest}, {Huang}, {Huffenberger},
  {Hurier}, {Jaffe}, {Jaffe}, {Jones}, {Juvela}, {Keih{\"a}nen}, {Keskitalo},
  {Kisner}, {Kneissl}, {Knoche}, {Knox}, {Kunz}, {Kurki-Suonio}, {Lagache},
  {L{\"a}hteenm{\"a}ki}, {Lamarre}, {Lasenby}, {Lattanzi}, {Lawrence}, {Leahy},
  {Leonardi}, {Lesgourgues}, {Levrier}, {Lewis}, {Liguori}, {Lilje},
  {Linden-V{\o}rnle}, {L{\'o}pez-Caniego}, {Lubin}, {Mac{\'\i}as-P{\'e}rez},
  {Maggio}, {Maino}, {Mandolesi}, {Mangilli}, {Marchini}, {Maris}, {Martin},
  {Martinelli}, {Mart{\'\i}nez-Gonz{\'a}lez}, {Masi}, {Matarrese}, {McGehee},
  {Meinhold}, {Melchiorri}, {Melin}, {Mendes}, {Mennella}, {Migliaccio},
  {Millea}, {Mitra}, {Miville-Desch{\^e}nes}, {Moneti}, {Montier}, {Morgante},
  {Mortlock}, {Moss}, {Munshi}, {Murphy}, {Naselsky}, {Nati}, {Natoli},
  {Netterfield}, {N{\o}rgaard-Nielsen}, {Noviello}, {Novikov}, {Novikov},
  {Oxborrow}, {Paci}, {Pagano}, {Pajot}, {Paladini}, {Paoletti}, {Partridge},
  {Pasian}, {Patanchon}, {Pearson}, {Perdereau}, {Perotto}, {Perrotta},
  {Pettorino}, {Piacentini}, {Piat}, {Pierpaoli}, {Pietrobon}, {Plaszczynski},
  {Pointecouteau}, {Polenta}, {Popa}, {Pratt}, {Pr{\'e}zeau}, {Prunet},
  {Puget}, {Rachen}, {Reach}, {Rebolo}, {Reinecke}, {Remazeilles}, {Renault},
  {Renzi}, {Ristorcelli}, {Rocha}, {Rosset}, {Rossetti}, {Roudier},
  {Rouill{\'e} d'Orfeuil}, {Rowan-Robinson}, {Rubi{\~n}o-Mart{\'\i}n},
  {Rusholme}, {Said}, {Salvatelli}, {Salvati}, {Sandri}, {Santos},
  {Savelainen}, {Savini}, {Scott}, {Seiffert}, {Serra}, {Shellard}, {Spencer},
  {Spinelli}, {Stolyarov}, {Stompor}, {Sudiwala}, {Sunyaev}, {Sutton},
  {Suur-Uski}, {Sygnet}, {Tauber}, {Terenzi}, {Toffolatti}, {Tomasi},
  {Tristram}, {Trombetti}, {Tucci}, {Tuovinen}, {T{\"u}rler}, {Umana},
  {Valenziano}, {Valiviita}, {Van Tent}, {Vielva}, {Villa}, {Wade}, {Wandelt},
  {Wehus}, {White}, {White}, {Wilkinson}, {Yvon}, {Zacchei}, \&
  {Zonca}}]{PlanckCollaboration2016}
{Planck Collaboration}, {Ade}, P.~A.~R., {Aghanim}, N., {et~al.} 2016, \aap,
  594, A13

\bibitem[{{Romani} {et~al.}(2004){Romani}, {Sowards-Emmerd}, {Greenhill}, \&
  {Michelson}}]{Romani2004}
{Romani}, R.~W., {Sowards-Emmerd}, D., {Greenhill}, L., \& {Michelson}, P.
  2004, \apjl, 610, L9

\bibitem[{{Saikia} {et~al.}(2001){Saikia}, {Jeyakumar}, {Salter}, {Thomasson},
  {Spencer}, \& {Mantovani}}]{Saikia2001}
{Saikia}, D.~J., {Jeyakumar}, S., {Salter}, C.~J., {et~al.} 2001, \mnras, 321,
  37

\bibitem[{{Shao} {et~al.}(2020){Shao}, {Wagg}, {Wang}, {Carilli}, {Riechers},
  {Intema}, {Weiss}, \& {Menten}}]{Shao2020}
{Shao}, Y., {Wagg}, J., {Wang}, R., {et~al.} 2020, \aap, 641, A85

\bibitem[{{Sharp} {et~al.}(2001){Sharp}, {McMahon}, {Irwin}, \&
  {Hodgkin}}]{Sharp2001}
{Sharp}, R.~G., {McMahon}, R.~G., {Irwin}, M.~J., \& {Hodgkin}, S.~T. 2001,
  \mnras, 326, L45

\bibitem[{{Slish}(1963)}]{Slish1963}
{Slish}, V.~I. 1963, \nat, 199, 682

\bibitem[{{Snellen} {et~al.}(1998){Snellen}, {Schilizzi}, {de Bruyn}, {Miley},
  {Rengelink}, {Roettgering}, \& {Bremer}}]{Snellen1998}
{Snellen}, I.~A.~G., {Schilizzi}, R.~T., {de Bruyn}, A.~G., {et~al.} 1998,
  \aaps, 131, 435

\bibitem[{{Snellen} {et~al.}(2000){Snellen}, {Schilizzi}, {Miley}, {de Bruyn},
  {Bremer}, \& {R{\"o}ttgering}}]{Snellen2000}
{Snellen}, I.~A.~G., {Schilizzi}, R.~T., {Miley}, G.~K., {et~al.} 2000, \mnras,
  319, 445

\bibitem[{{Spingola} {et~al.}(2020){Spingola}, {Dallacasa}, {Belladitta},
  {Caccianiga}, {Giroletti}, {Moretti}, \& {Orienti}}]{Spingola2020}
{Spingola}, C., {Dallacasa}, D., {Belladitta}, S., {et~al.} 2020, \aap, 643,
  L12

\bibitem[{{Stanghellini} {et~al.}(1998){Stanghellini}, {O'Dea}, {Dallacasa},
  {Baum}, {Fanti}, \& {Fanti}}]{Stanghellini1998}
{Stanghellini}, C., {O'Dea}, C.~P., {Dallacasa}, D., {et~al.} 1998, \aaps, 131,
  303

\bibitem[{{Stern} {et~al.}(2003){Stern}, {Hall}, {Barrientos}, {Bunker},
  {Elston}, {Ledlow}, {Raines}, \& {Willis}}]{Stern2003}
{Stern}, D., {Hall}, P.~B., {Barrientos}, L.~F., {et~al.} 2003, \apjl, 596, L39

\bibitem[{{Stocke} {et~al.}(1992){Stocke}, {Morris}, {Weymann}, \&
  {Foltz}}]{Stocke1992}
{Stocke}, J.~T., {Morris}, S.~L., {Weymann}, R.~J., \& {Foltz}, C.~B. 1992,
  \apj, 396, 487

\bibitem[{{Thomasson} {et~al.}(2003){Thomasson}, {Saikia}, \&
  {Muxlow}}]{Thomasson2003}
{Thomasson}, P., {Saikia}, D.~J., \& {Muxlow}, T.~W.~B. 2003, \mnras, 341, 91

\bibitem[{{Tingay} \& {de Kool}(2003)}]{Tingay2003}
{Tingay}, S.~J. \& {de Kool}, M. 2003, \aj, 126, 723

\bibitem[{{Tingay} {et~al.}(2015){Tingay}, {Macquart}, {Collier}, {Rees},
  {Callingham}, {Stevens}, {Carretti}, {Wayth}, {Wong}, {Trott}, {McKinley},
  {Bernardi}, {Bowman}, {Briggs}, {Cappallo}, {Corey}, {Deshpande}, {Emrich},
  {Gaensler}, {Goeke}, {Greenhill}, {Hazelton}, {Johnston-Hollitt}, {Kaplan},
  {Kasper}, {Kratzenberg}, {Lonsdale}, {Lynch}, {McWhirter}, {Mitchell},
  {Morales}, {Morgan}, {Oberoi}, {Ord}, {Prabu}, {Rogers}, {Roshi}, {Udaya
  Shankar}, {Srivani}, {Subrahmanyan}, {Waterson}, {Webster}, {Whitney},
  {Williams}, \& {Williams}}]{Tingay2015}
{Tingay}, S.~J., {Macquart}, J.~P., {Collier}, J.~D., {et~al.} 2015, \aj, 149,
  74

\bibitem[{{Tinti} {et~al.}(2005){Tinti}, {Dallacasa}, {de Zotti}, {Celotti}, \&
  {Stanghellini}}]{Tinti2005}
{Tinti}, S., {Dallacasa}, D., {de Zotti}, G., {Celotti}, A., \& {Stanghellini},
  C. 2005, \aap, 432, 31

\bibitem[{{Torniainen} {et~al.}(2005){Torniainen}, {Tornikoski},
  {Ter{\"a}sranta}, {Aller}, \& {Aller}}]{Torniainen2005}
{Torniainen}, I., {Tornikoski}, M., {Ter{\"a}sranta}, H., {Aller}, M.~F., \&
  {Aller}, H.~D. 2005, \aap, 435, 839

\bibitem[{{van Breugel} {et~al.}(1984){van Breugel}, {Miley}, \&
  {Heckman}}]{vanBreugel1984}
{van Breugel}, W., {Miley}, G., \& {Heckman}, T. 1984, \aj, 89, 5

\bibitem[{{Wiita}(2004)}]{Wiita2004}
{Wiita}, P.~J. 2004, \apss, 293, 235

\bibitem[{{Williams} {et~al.}(2016){Williams}, {van Weeren}, {R{\"o}ttgering},
  {Best}, {Dijkema}, {de Gasperin}, {Hardcastle}, {Heald}, {Prandoni},
  {Sabater}, {Shimwell}, {Tasse}, {van Bemmel}, {Br{\"u}ggen}, {Brunetti},
  {Conway}, {En{\ss}lin}, {Engels}, {Falcke}, {Ferrari}, {Haverkorn},
  {Jackson}, {Jarvis}, {Kapi{\'n}ska}, {Mahony}, {Miley}, {Morabito},
  {Morganti}, {Orr{\'u}}, {Retana-Montenegro}, {Sridhar}, {Toribio}, {White},
  {Wise}, \& {Zwart}}]{Williams2016}
{Williams}, W.~L., {van Weeren}, R.~J., {R{\"o}ttgering}, H.~J.~A., {et~al.}
  2016, \mnras, 460, 2385

\bibitem[{{Willott} {et~al.}(2010){Willott}, {Delorme}, {Reyl{\'e}}, {Albert},
  {Bergeron}, {Crampton}, {Delfosse}, {Forveille}, {Hutchings}, {McLure},
  {Omont}, \& {Schade}}]{Willott2010}
{Willott}, C.~J., {Delorme}, P., {Reyl{\'e}}, C., {et~al.} 2010, \aj, 139, 906

\bibitem[{{Wolf} {et~al.}(2021){Wolf}, {Nandra}, {Salvato}, {Liu}, {Buchner},
  {Brusa}, {Hoang}, {Moss}, {Arcodia}, {Br{\"u}ggen}, {Comparat}, {de
  Gasperin}, {Georgakakis}, {Hotan}, {Lamer}, {Merloni}, {Rau}, {Rottgering},
  {Shimwell}, {Urrutia}, {Whiting}, \& {Williams}}]{Wolf2021}
{Wolf}, J., {Nandra}, K., {Salvato}, M., {et~al.} 2021, \aap, 647, A5

\bibitem[{{Yang} {et~al.}(2021){Yang}, {Urquhart}, {Thompson}, {Menten},
  {Wyrowski}, {Brunthaler}, {Tian}, {Rugel}, {Yang}, {Yao}, \&
  {Mutale}}]{Yang2021}
{Yang}, A.~Y., {Urquhart}, J.~S., {Thompson}, M.~A., {et~al.} 2021, \aap, 645,
  A110

\bibitem[{{Yi} {et~al.}(2014){Yi}, {Wang}, {Wu}, {Yang}, {Bai}, {Fan},
  {Brandt}, {Ho}, {Zuo}, {Kim}, {Wang}, {Yang}, {Zhang}, {Wang}, {Wang}, {Ai},
  {Fan}, {Chang}, {Wang}, {Lun}, \& {Xin}}]{Yi2014}
{Yi}, W.-M., {Wang}, F., {Wu}, X.-B., {et~al.} 2014, \apjl, 795, L29

\bibitem[{{Zeimann} {et~al.}(2011){Zeimann}, {White}, {Becker}, {Hodge},
  {Stanford}, \& {Richards}}]{Zeimann2011}
{Zeimann}, G.~R., {White}, R.~L., {Becker}, R.~H., {et~al.} 2011, \apj, 736, 57

\bibitem[{{Zovaro} {et~al.}(2019){Zovaro}, {Sharp}, {Nesvadba}, {Bicknell},
  {Mukherjee}, {Wagner}, {Groves}, \& {Krishna}}]{Zovaro2019}
{Zovaro}, H. R.~M., {Sharp}, R., {Nesvadba}, N. P.~H., {et~al.} 2019, \mnras,
  484, 3393

\end{thebibliography}

\begin{appendix}

\section{Radio continuum images}
\label{images}
The ten quasars at $z>5$ in our new VLA and uGMRT observations are all point sources. Below we present their radio continuum images.

\begin{figure*}[h]
\centering
\subfigure{\includegraphics[scale=0.069]{vla1.pdf}} 
\caption{Continuum images from (left to right) the VLA S, C, and X bands. The three panels in each row present the VLA three-band images for one target. Each panel has an image size of $3\arcsec\times3\arcsec$. The black crosses mark the published optical positions (\citealt{Sharp2001}; \citealt{McGreer2009}; SDSS); in the case of J0131$-$0321, J1146+4037, and J2228+0110 we adopt the position from EVN observations (\citealt{Frey2010}; \citealt{Cao2014}; \citealt{Gabanyi2015}). The shapes of the synthesized beams are plotted in the bottom left corner of each sub-figure. The beam sizes are presented in Table \ref{vlaobs}, where the flux density for each target is also listed. The contour levels are [$-$4, 4, 16, 64, 256, 1024] $\times$ rms (the 1$\sigma$ off-source noise level listed in Table \ref{vlaobs}).} 
\label{vlaima}
\end{figure*}

\begin{figure*}
\centering
\subfigure{\includegraphics[scale=0.069]{vla2.pdf}} 
\leftline{Fig. \ref{vlaima}. Continued.}
\end{figure*}

\begin{figure*}
\centering
\subfigure{\includegraphics[scale=0.077]{ugmrt.pdf}} 
\caption{Continuum images from uGMRT. The top left panel shows the band-3 image of J0131$-$0321, and the rest are band-4 images. Each panel has an image size of $30\arcsec\times30\arcsec$. The black crosses are the published optical positions in  \citet{McGreer2009} and SDSS; in the case of J0131$-$0321, J0836+0054, J0913+5919, J1146+4037, and J1427+3312, we use the positions from VLBA or EVN observations (\citealt{Frey2003, Frey2010}; \citealt{Momjian2003, Momjian2008}; \citealt{Gabanyi2015}). The shapes of the synthesized beams are plotted in the bottom left corner of each sub-figure. The beam size of each map and the flux density of each target are presented in Table \ref{ugmrtobs}. The contour levels are [$-$4, 4, 8, 16, 32, 64] $\times$ rms (the 1$\sigma$ off-source noise level listed in Table \ref{ugmrtobs}).}
\label{ugmrtima}
\end{figure*}

\section{Model-fitting results for individual sources}
\label{notes}
We present the model-fitting results for each target as follows.

{\bf J0131$-$0321 - }The radio spectrum toward this target peaks at $\sim$10 GHz in the rest frame ($\sim$2 GHz in the observed frame). Our new VLA S-, C-, and X-band observations (black points with error bars in Fig. \ref{nine}) describe the frequency range above the peak frequency, which present an optically thin power-law index of $\sim-0.7$.  
In the region below the peak frequency, we also detected this target with uGMRT bands 3 and 4 (black open diamonds  with error bars in Fig. \ref{nine}). These together with the VLA FIRST  1.4 GHz \citep{Helfand2015} and 1.7 GHz \citep{Gabanyi2015}  measurements, make the 150 MHz measurement from the TGSS \citep{Intema2017} database an outlier. Considering that our new uGMRT band-3 and band-4 observations were taken on adjacent dates (2020 October 01 and 04), however, the 150 MHz data (with a resolution of 25$\arcsec$) were observed on 2016 March 15, there may be strong variability during the $\sim$4 years for this target in the low-frequency (e.g., observing frequency below 1 GHz) part. However, \citet{Hajela2019} conducted a dedicated transient survey of 300 deg$^{2}$ of the SDSS Stripe 82 region using the GMRT at 150 MHz, and they find the sky at low frequency shows very little variability (i.e., only 0.05$\%$ is variable on a timescale of 4 years). Further multi-epoch 150 GHz monitoring of this target is necessary. The possibility of contamination by any bright radio sources near J0131$-$0321 can be removed by comparing the flux density at observed frame 1.4 GHz by FIRST (33.69$\pm$0.12 mJy with resolution of 5$\arcsec$) and  by the NRAO VLA Sky Survey (31.4$\pm$1.0 mJy with resolution of 45$\arcsec$). 
 As shown in Fig. \ref{nine}, neither homogeneous SSA nor homogeneous FFA models  can accurately describe the spectrum, which can better fit the frequency above $\sim$10 GHz in the rest frame but overestimate most data points at the low-frequency end. Both internal FFA and inhomogeneous FFA models can well describe the entire radio spectrum, with the maximum likelihood value of the best-fit inhomogeneous FFA model slightly higher than that of internal FFA model.

{\bf J0741+2520 - }The radio spectrum of this target peaks at $\sim$20 GHz in the rest frame ($\sim$3 GHz in the observed frame). The data points from the continuous spectral windows from our new VLA S-band (frequency coverage: 2--4 GHz) observations are distributed on both sides of the spectral peak. The VLA S-band data themselves show evidence of the spectral curvature. In the frequency range above the peak frequency,  our new VLA S-, C-, and X-band detections present a steep spectrum with optically thin power-law index of $\sim$--0.7. In the frequency range below the peak frequency, we detected this target in uGMRT band 4, together with our new VLA S-band data and archival data (\citealt{Helfand2015}; \citealt{Shao2020}), can well constrain the gradient of the optically thick part. As shown in Fig. \ref{nine}, the homogeneous SSA, homogeneous FFA, and internal FFA models seem to fail to mimic the spectrum with frequency range below the peaked frequency. The inhomogeneous FFA model can better fit the entire radio spectrum compared with the other three ones, with fitting parameters shown in Table \ref{fitr}.

{\bf J0836+0054 - }The radio spectral turnover of this target is at $\sim$6 GHz in the rest frame ($\sim$0.9 GHz in the observed frame). 
In the frequency range above the turnover frequency, the archival data from observed frame 1.4 GHz to 5 GHz (\citealt{Frey2003, Frey2005}; \citealt{Petric2003}) constrain the optically thin power-law index to be $\sim$--1.2. 
In the frequency range below the turnover frequency, our new uGMRT band-4 detection together with archival data (\citealt{Shao2020}; \citealt{Wolf2021}) present a very flat spectrum with a spectral index close to 0.  As shown in Fig. \ref{nine}, the homogeneous SSA, homogeneous FFA, and internal FFA models are not able to describe a flat spectrum in the optically thick part. But the inhomogeneous FFA model can  fit the entire radio spectrum toward this target.

{\bf J0913+5919 - } As shown in our previous paper \citep{Shao2020}, this target may have decreasing flux density with increasing frequency. Our new uGMRT band-4 detection, together with archival data (\citealt{Momjian2003}; \citealt{Petric2003}; \citealt{Shao2020}), show evidence of a spectral peak. This makes the data point observed at 232 MHz by GMRT with a flux density of $S_{\rm 232MHz}=30.0\pm3.0$ mJy \citep{Carilli2007}, we previously used in \citep{Shao2020}, an outlier. \citet{Coppejans2017} re-reduced the 232 MHz data and reported  strong image-plane ripples in the central region
near the source, which is likely the result of baseline-based errors that were a  common feature in older GMRT data and affect flux density measurements. They measure a flux density of  $10.7\pm1.2$ mJy, which we adopted in this work. The radio spectrum of this target peaks at $\sim$20 GHz in the rest frame. As shown in Fig. \ref{nine}, the homogeneous SSA and homogeneous FFA models cannot accurately fit the low-frequency data. The internal FFA and inhomogeneous FFA model can better fit the entire radio spectrum with the maximum likelihood value of the best-fit internal FFA model slightly higher than that of the inhomogeneous FFA model. We present the best-fit parameters in Table \ref{fitr}.

{\bf J1034+2033 - }The radio spectrum of this target turns over at $\sim$20 GHz in the rest frame.  Our new VLA S-, C-, and X-band observations sample above the turnover frequency, which reveal an optically thin power-law index of $\sim$--1.0.  In the region below the turnover frequency, we also detect this target with uGMRT band 4. Together with  VLA FIRST 1.4 GHz \citep{Helfand2015} and GMRT 323 MHz \citep{Shao2020} measurements, the optically thick part shows a rather flat spectrum. As shown in Fig. \ref{nine}, the homogeneous SSA, homogeneous FFA, and internal FFA models cannot accurately describe the optically thick  flat spectrum. However, the inhomogeneous FFA model is able to describe the entire radio spectrum toward this target.

{\bf J1146+4037 - }
The radio spectrum of J1146+4037 shows evidence of two spectral turnovers, one at $\sim$10 GHz in the rest frame and the other at  $\sim$50 GHz in the rest frame. 
Our new VLA S-band detections indicate a steep synchrotron optically thin spectrum, and further to the lower frequency in the optically thick part,  our new uGMRT band-4 detection and archival data (\citealt{Frey2010}; \citealt{Helfand2015}; \citealt{Shao2020}) present a prominent spectral turnover.
Our new VLA C- and X-band observations together reveal an additional weak peak.
This complex radio spectrum may be caused  by multiple radio components, as discussed in Sect. \ref{cs}. We modeled the entire radio spectrum with  two same absorption models and two different absorption models, as shown in Fig. \ref{J1146modelfit} (for the latter case, we only present the best-fit ones). As for the same absorption models case, double inhomogeneous FFA models can best fit the entire radio spectrum. The best-fit model shows similar optically thin power-law index of $\sim$--1.8 for both components as illustrated in Table \ref{fitrJ1146}. As for the different absorption models cases, inhomogeneous FFA+homogeneous SSA and internal FFA+homogeneous SSA models can accurately describe the observed radio spectrum.

{\bf J1427+3312 - }The radio spectrum of J1427+3312 reveals evidence for a  turnover at $\sim$300 MHz in the observed frame ($\sim$2 GHz in the rest frame).
We detected this target in uGMRT band 4. Together with archival data (\citealt{Frey2008}; \citealt{Momjian2008}; \citealt{Coppejans2015}; \citealt{Shao2020}), this target shows a steep power-law spectrum with index of $\sim$--0.9. Moving to the lower-frequency region, \citet{Williams2016} present the GMRT 150 MHz flux density below the spectral trend mentioned above. This may indicate a spectral turnover around $\sim$300 MHz in the observed frame. However, as our uGMRT band-2 observations toward this target are unusable  due to severe RFI, the optically thick part spectrum is less constrained only by one 150 MHz data point \citep{Williams2016}. Thus, as shown in Fig. \ref{nine} and Table \ref{fitr}, the homogeneous SSA, homogeneous FFA , the internal FFA and the inhomogeneous FFA models can all well describe the observed radio spectrum. From the fitting aspect, the maximum likelihood value of the best-fit homogeneous  FFA model slightly higher than that of other three models.

{\bf J1614+4640 - }The radio spectrum of this target peaks at $\sim$25 GHz in the rest frame ($\sim$4 GHz in the observed frame). The spectral peak lies between the VLA S band (frequency coverage: 2--4 GHz) and C band (frequency coverage: 4--8 GHz). In the frequency range above the peak frequency,  our new VLA C- and X-band detections constrain a steep spectrum with an optically thin power-law index of $\sim$--1.2. In the frequency range below the peak frequency,  our new VLA S-band  and uGMRT band-4 detections, and archival data (\citealt{Helfand2015}; \citealt{Shao2020}) present a steep spectrum in the optically thick part. As shown in Fig. \ref{nine}, the homogeneous SSA and homogeneous FFA models underestimate the flux density below observed frame 1 GHz. Both the internal FFA and inhomogeneous FFA models successfully fit all the data points in the radio regime.

{\bf J2228+0110 - }This target is the weakest radio source in our sample. We detect this target in all VLA bands, but with low S/N (e.g., 5--10). Thus, we imaged the data using all spectral windows for each observing band. Our new VLA observations, together with archival data (\citealt{Hodge2011}; \citealt{Cao2014}; \citealt{Shao2020}), reveal a power-law-like radio spectrum, ranging from 323 MHz to 10 GHz in the observed frame, or ranging from 1.4 GHz to 10 GHz only considering the detections. During the model fitting, we only considered the detections, which results in  a flat spectrum with a power-law index of $-0.39^{+0.16}_{-0.17}$. The best fitting power-law model predicts a flux density larger than the 3$\sigma$ upper limit at 323 MHz (\citealt{Shao2020}; see Fig. \ref{nine}). This may reveal a spectral turnover at approximately a few gigahertz in the rest frame. High sensitivity low-frequency observations with, for example, the uGMRT are needed to check this.
The VLA S-, C-, and X-band observations alone show a bump-like feature with the flux density at the C band being larger than the flux densities measured at  S  and X bands, even with large uncertainties. This may be due to source variability as the observations of the three bands are separated by about 22 days.

{\bf J2245+0024 - }The radio spectrum of this target peaks at $\sim$30 GHz in the rest frame ($\sim$5 GHz in the observed frame). The spectral peak lies in VLA C band (frequency coverage: 4--8 GHz). In the frequency range above the peak frequency,  our new VLA C- and X-band detections constrain the optically thin power-law index of $\sim$--0.7. In the frequency range below the peak frequency,  our new VLA S- and C-band observations and the 1.4 GHz archival data from VLA high-resolution radio survey of SDSS strip 82 \citep{Hodge2011} can constrain the optically thick part spectrum. As shown in Fig. \ref{nine}, the homogeneous SSA, homogeneous FFA, and internal FFA models cannot well fit the spectral curvature showing by the VLA C-band data. The best-fit inhomogeneous FFA model predicts a slightly higher flux density at observed frame 323 MHz than the 3$\sigma$ upper limit presented by \citet{Shao2020}, but they are consistent within the fitting error showing by the shadow region in Fig. \ref{nine}.

\end{appendix}

\end{document}